\documentclass[preprint2,,trackchanges]{aastex6}

\newcommand{\pcc}{\mathrm{cm}^{-3}}

\newcommand{\Pturb}{\langle P_\mathrm{ram}\rangle}
\newcommand{\Pmag}{\langle \Pi_\mathrm{mag}\rangle}
\newcommand{\Pth}{\langle P_\mathrm{th}\rangle}
\newcommand{\Ptot}{\langle P_\mathrm{tot}\rangle}
\newcommand{\nave}{\langle n_0\rangle}
\newcommand{\nsh}{\langle n_\mathrm{sh}\rangle}
\newcommand{\rhosh}{\langle \rho_\mathrm{sh}\rangle}
\newcommand{\rhoave}{\langle \rho_0\rangle}
\newcommand{\CAp}{C_\mathrm{A0\perp}}

\newcommand{\Bshp}{B_\mathrm{sh\perp}}

\newcommand{\dvxeq}{\delta v_{x,\mathrm{eq}}}
\newcommand{\thetacr}{\theta_{\mathrm{cr}}}

\newcommand{\Bcr}{B_{\mathrm{cr}}}

\fullcollaborationName{The Friends of AASTeX Collaboration}

\begin{document}


\title{
The Early Stage of Molecular Cloud Formation by Compression of Two-phase Atomic Gases
}


\author{
Kazunari Iwasaki\altaffilmark{1}, 
Kengo Tomida\altaffilmark{1},
Tsuyoshi Inoue\altaffilmark{2}, 
and Shu-ichiro Inutsuka\altaffilmark{2}
}


\altaffiltext{1}{
Department of Earth and Space Science, Osaka University, 
Machikaneyama-cho, Toyonaka, Osaka 560-0043, Japan,\\
iwasaki@astro-osaka.jp
}
\altaffiltext{2}{
Department of Physics, Nagoya University, 
Furo-cho, Chikusa-ku, Nagoya, Aichi 464-8602, Japan
}

\begin{abstract}
We investigate the formation of molecular clouds from atomic gas by using 
three-dimensional magnetohydrodynamic simulations, 
including non-equilibrium chemical reactions and heating/cooling processes.
We consider super-Alfv\'enic head-on colliding flows of atomic gas 
possessing the two-phase structure that
consists of HI clouds and surrounding warm diffuse gas.
We examine how the formation of molecular clouds depends
on the angle $\theta$ between the upstream flow and the mean magnetic field. 
We find that 
there is a critical angle $\theta_\mathrm{cr}$ above which the shock-amplified magnetic field 
controls the post-shock gas dynamics.
If the atomic gas is compressed almost along the mean magnetic field ($\theta\ll\theta_\mathrm{cr}$), 
super-Alfv\'enic anisotropic turbulence is maintained
by the accretion of the highly inhomogeneous upstream atomic gas.
As a result, a greatly extended turbulence-dominated post-shock layer is generated.
Around $\theta\sim \theta_\mathrm{cr}$, 
the shock-amplified magnetic field weakens the post-shock turbulence, leading to 
a dense post-shock layer. 
For $\theta\gg \theta_\mathrm{cr}$, the strong magnetic pressure suppresses the formation of 
cold dense clouds.
Efficient molecular cloud formation is expected if $\theta$ is less than a few times 
$\theta_\mathrm{cr}$.
Developing an analytic model and performing a parameter survey, we obtain 
an analytic formula for the critical angle as a function of the mean density, collision speed, and 
field strength of the upstream atomic gas. 
The critical angle is found to be less than $\sim 15^\circ$ 
as long as the field strength is larger than $1~\mu$G, indicating that  
the probability of occurrence of compression with $\theta<\theta_\mathrm{cr}$ is limited if
shock waves come from various directions.

\end{abstract}

\keywords{ISM --- ISM:clouds -- ISM:magnetic fields -- ISM:molecules -- stars:formation}

\section{Introduction} \label{sec:intro}
The formation of molecular clouds (MCs) is one of the fundamental building blocks 
in star formation.
MCs are thought to form through shock compression of atomic gas.
In galaxies, shock waves are frequently driven by several energetic phenomena, including
supernova (SN) explosions \citep{Chevalier1974,MO1977,KI2000}, 
expansion of super-bubbles due to multiple SNe 
\citep{McCray1987,Tomisaka1988,Dawson2011,Dawson2013a,Ntormousi2017}, 
and galactic spiral waves \citep{Wada2011,Grand2012,Baba2017}.

Head-on colliding flows of warm neutral medium (WNM) have been considered to model the formation of MCs.
The simple colliding flow model allows us to investigate the detailed physics 
working in the shocked regions using local numerical simulations with high resolution.
Although the setup is simple, 
the shock compression generates a complex turbulent multi-phase structure consisting of 
shocked WNM and cold clouds condensed through the thermal instability 
\citep{KI2002,AH2005,Heitsch2005,Heitsch2006,Vaz2006,
Vaz2007,HA2007,HAM07,Henn2008,Banerjee2009,Heitsch2009,AH2010,Vaz2010,Vaz2011,Clark2012}.
The turbulence is driven by various instabilities \citep{Heitsch2008}, including
the thermal instability 
\citep{Field1965,Balbus1986,KI2000,IT2008,IT2009}, 
the Kelvin-Helmholtz and Rayleigh-Taylor instabilities \citep{Cha1961}, and
the nonlinear
thin-shell instability \citep{Vis1994,Blondin1996,Heitsch2007,Folini2014}.
These simulation results successfully
reproduce observational properties of the interstellar clouds \citep{HAM07}, such as 
the supersonic turbulence \citep{Larson1979,Larson1981,Solomon1987,Heyer2009}, 
the mass function of cold clumps \citep{Kramer1998,Heithausen1998}, and 
the angular correlation between filamentary atomic gases and magnetic fields 
\citep{Clark2015,Inoue2016}.

Magnetic fields exert a great impact on the MC formation.
Performing magnetohydrodynamic (MHD) simulations,
\citet{II2008,II2009} showed that interstellar clouds, which are the precursor of MCs,
are formed only if the WNM is compressed almost along the magnetic field.
Otherwise, the formation of dense clouds is prohibited by the shock-amplified magnetic field, 
and only HI clouds form \citep[also see][]{Henn2000,Heitsch2009,vanLoo2010,Kortgen2015}.
If the WNM is compressed by shock waves 
from various directions with respect to magnetic fields, 
the MC formation is expected to be significantly delayed and HI clouds are formed in almost all 
compression events.
Most simulations of the MC formation have focused on colliding flows parallel to the magnetic field
\citep{Henn2008,Banerjee2009,Vaz2011,Valdivia2016,Zam2018}.

Although most authors have considered the MC formation directly from a typical atomic gas 
with a density of $\sim 1~\pcc$, 
there is some observational evidence that MCs grow through accretion of HI clouds 
with a density of $\sim 10~\pcc$
\citep{Blitz2007,Fukui2009,Fukui2017} and 
even through accretion of denser clouds containing molecules \citep{Motte2014}.
In addition, recent global simulations of galaxies revealed that 
the direct precursor of MCs is not typical diffuse atomic gases but dense atomic gases
with a mean density of $\sim 20~\mathrm{cm}^{-3}$
which have been piled up by dynamical disturbances \citep{Dobbs2012,Bonnell2013,Baba2017}.
The formation of molecules requires 
a column density larger than $\sim 2\times 10^{21}~\mathrm{cm}^{-2}$ \citep{vanDishoeck1988}.
The accumulation length to form molecules is required to be larger than $650~\mathrm{pc}~(n/1~\pcc)^{-1}$,
where $n$ is the gas density \citep{Hartmann2001,Pringle2001,McKee2007,II2009}.
This length may be too long to form MCs within a typical lifetime of a few tens of megayears 
estimated from observational results \citep{Kawamura2009,Murray2011,Meidt2015}.
From the observational and theoretical points of view mentioned above, 
shock compression of gases denser than a typical atomic gas with a density of $\sim 1~\pcc$ is 
one plausible path from atomic gas to MCs.
The dense atomic gases are generated by multiple episodes of shock compression 
\citep{II2009,Inutsuka2015,Kobayashi2017,Kobayashi2018}.

\citet{Inoue2012} investigated the MC formation 
directly from collision of a dense atomic gas 
with a mean density of $\sim 5~\mathrm{cm}^{-3}$ along the magnetic field.
They took into account the two-phase structure of the atomic gas 
\citep{FGH1969,Wetal95,Wetal03,II2014}.
They found that the collision of the highly inhomogeneous upstream gas 
drives super-Alfv{\'e}nic anisotropic turbulence. 
MCs form on a timescale of $10$~Myr which 
is consistent with that estimated observationally by \citet{Kawamura2009}.

Shock compression completely along the magnetic field considered in \citet{Inoue2012} is 
an extreme case. 
If the atomic gas is compressed from various directions randomly, 
shock compression misaligned to magnetic fields is more likely to occur.
Although \citet{Henn2000} and \citet{II2009} have investigated the effect of 
field orientation on the cloud formation, their studies are restricted to the cases with colliding flows 
of the WNM with a density less than $1~\pcc$.
Atomic gas intrinsically has the two-phase structure, and 
the resulting density inhomogeneity is expected to significantly affect the MC formation. 
In this paper, taking into account the realistic two-phase structure of atomic gas, 
we clarify how the MC formation depends on the angle 
between the upstream flow and magnetic field 
by performing three-dimensional magnetohydrodyamic simulations including 
simplified chemical reactions and cooling/heating processes.

This paper is organized as follows.
Section \ref{sec:method} presents the basic equations, and methods for chemical reactions, 
heating/cooling processes, and simplified ray tracing.
In Section \ref{sec:results}, the results of a fiducial parameter set are presented.
In Section \ref{sec:ana}, we introduce an analytic model that describes the simulation results.
The results of a parameter survey are shown in Section \ref{sec:parameter}.
Astrophysical implications are discussed in Section \ref{sec:discussion}.
Finally, our results are summarized in Section \ref{sec:summary}.

\section{Equations and Methods}\label{sec:method}

\subsection{Basic Equations}
The basic equations are given by 
\begin{equation}
\frac{\partial \rho}{\partial t} + \frac{\partial \left( \rho v_\mu \right)  }{\partial x_\mu}= 0,
\label{eoc}
\end{equation}
\begin{equation}
\frac{\partial \rho v_\mu}{\partial t} 
+ \frac{\partial }{\partial x_\nu} \left( \rho v_\mu v_\nu + T_{\mu\nu} \right) = 0,
\end{equation}
\begin{equation}
\frac{\partial E}{\partial t} 
+ \frac{\partial }{\partial x_\mu} 
\left[ \left(E\delta_{\mu\nu} + T_{\mu\nu} \right) v_\nu  
- \kappa \frac{\partial T}{\partial x_\mu}\right] = -{\cal L},
\end{equation}
\begin{equation}
\frac{\partial B_\mu}{\partial t} 
+ \frac{\partial }{\partial x_\nu} \left(  v_\nu B_\mu - v_\mu B_\nu \right)=0,
\label{induc}
\end{equation}
and
\begin{equation}
     T_{\mu\nu} = \left( P + \frac{B^2}{8\pi} \right)\delta_{\mu\nu}  - \frac{B_\mu B_\nu}{4\pi},
     \label{Tij}
\end{equation}
where $\rho$ is the gas density, $v_\mu$ is the gas velocity, $B_\mu$ is the magnetic field, 
$E=\rho v^2/2 + P/(\gamma-1) + B^2/8\pi$ is the total energy density, 
$T$ is the gas temperature, 
$\kappa$ is the thermal conductivity,
and ${\cal L}$ is a net cooling rate per unit volume.
The thermal conductivity for neutral hydrogen is given by 
$\kappa(T) = 2.5\times 10^3~T^{1/2}$~cm$^{-1}$~K$^{-1}$~s$^{-1}$ \citep{Parker1953}.

Anisotropy of the thermal conduction is not important in our simulations 
because the ISM considered in this paper is weakly ionized.
In the weakly ionized ISM, the main carriers of heat are neutral particles. 
The threshold temperature below which the effect of charged particles on the thermal conduction 
coefficient
is negligible is approximately $5\times 10^4~\mathrm{K}$ (Parker 1953).
In our simulations, the gas temperature increases to 
$\sim 10^4$~K through the shock compression and 
decreases due to radiative cooling, indicating that the gas temperature is lower than 
the threshold temperature throughout the computation box.

To solve the basic equations (\ref{eoc})-(\ref{induc}), 
we use Athena++ \citep[][in preparation]{Stone2018} 
which is a complete
rewrite of the Athena MHD code \citep{Stone2008}. 
The HLLD method is used as the MHD Riemann solver \citep{MK2005}.
Magnetic fields are integrated with the constrained transport method 
\citep{EH1988,Gardiner2008}.

Chemical reactions, heating/cooling processes, 
and ray tracing of far-ultraviolet (FUV) photons and cooling photons
are newly implemented into Athena++ as shown in the next section.
Because self-gravity is ignored  in this paper, we focus on the early phase of the MC formation.
The effect of self-gravity will be investigated in forthcoming papers.

\subsection{Chemical Reactions and Heating/Cooling Processes}

To treat the transition from atomic to molecular phases, 
non-equilibrium chemical reactions must be taken into account \citep{Glover2007,Clark2012,Valdivia2016}.
We follow \citet{Inoue2012} for the chemical reactions, 
heating/cooling processes, and ray tracing of FUV and cooling photons.
We reduce the number of chemical species, retaining H$^+$, H, H$_2$, He$^+$, He, and 
the most important species for cooling, C$^+$, O, and CO.
The elemental abundances of He, C, and O with respect to hydrogen nuclei are 
given by ${\cal A}_\mathrm{He}=0.1$, ${\cal A}_\mathrm{C}=1.5\times 10^{-4}$, and 
${\cal A}_\mathrm{O}=3.2\times 10^{-4}$, respectively.
In the formation of CO, we use a simplified treatment of the conversion of C$^+$ to CO proposed by 
\citet{NL1997}\footnote{
Note that the evolutionary equation for CO in \citet{NL1997} has a typographical error 
pointed out in \citet{GC2012} although the error does not significantly affect the results.
In this paper, we use the corrected version of that expression.
}. 
\citet{GC2012} showed that gas dynamics is not sensitive to the detailed chemistry, and
the simplified method works reasonably well to capture the global behavior of the CO formation,
although CO abundance is not very accurate.
The FUV radiation field strength is set to $1.6\times 10^{-3}$~erg~cm$^{-2}$~s$^{-1}$,
corresponding to $G_0=1$ in the units of the Habing flux
\citep{Habing1968,Draine1978}.

The detailed chemical reactions and cooling/heating processes are described in \citet{Inoue2012}.
One update
is that the OI cooling is evaluated by using calculations of
the populations of the three levels exactly.

The number density of the $i$th species, $n_i$, satisfies the following equation:
\begin{equation}
\frac{\partial n_i}{\partial t} + \frac{\partial \left( n_i v_\mu \right)  }{\partial x_\mu}
= F_i - D_i,
\label{rate_eq}
\end{equation}
where $F_i$ and $D_i$ are the formation and destruction rates of the $i$th species.
The number density of electron is derived using
charge neutrality, $n(e) = n(\mathrm{H}^+) + n(\mathrm{He}^+)
+ n(\mathrm{C}^+)$.
In this paper, the number density is referred to as that of hydrogen 
nuclei, $n = n(\mathrm{H}) + n(\mathrm{H}^+) + 2n(\mathrm{H_2})$.

The density is given by 
\begin{equation}
     \rho = \sum_i m_i n_i  = \mu m_\mathrm{p} n,
\end{equation}
where $m_i$ is the mass of the $i$th species, $m_\mathrm{p}$ is the proton mass, and 
$\mu = 1.4$ is the mean molecular weight.
The equation of state is given by 
\begin{equation}
	P = n_\mathrm{tot} k_\mathrm{B} T,
     \label{eos}
\end{equation}
where $n_\mathrm{tot}$ is the total number density of the gas particles, 
$n_\mathrm{tot} = 
\left( 1 + {\cal A}_\mathrm{He} + {\cal A}_\mathrm{C} + {\cal A}_\mathrm{O}\right) n 
+ n(e) - n(\mathrm{H}_2) - n(\mathrm{CO})$.

Since the timescale of the chemical reactions is much shorter than the dynamical 
and cooling timescales,
a special treatment is required for solving the chemical reactions.
We divide Equation (\ref{rate_eq}) into the advection term and chemical reactions in an
operation-splitting manner. 
First, we solve the advection of the chemical species coupled with equations 
(\ref{eoc})-(\ref{induc}).
In order to keep consistency between equations (\ref{eoc}) and (\ref{rate_eq}) and 
to ensure the local conservation of the elemental abundances, we use the consistent 
multi-fluid advection algorithm \citep{PM1999,Glover2010}.
After that the chemical reactions are solved. 
\citet{II2008} and \citet{Inoue2012} developed the piecewise exact solution (PES) method.
The PES method divides the chemical reactions into small sets of reactions, each of 
which has an analytic solution, and solves them in an operator-splitting manner.
The advantage of the PES method is that iteration is not required.
Comparing the results obtained using the PES method and an iterative 
implicit method with the Gauss-Seidel method,
the difference is negligible at least in our simplified chemical network.
Since the PES method takes a shorter calculation time than the iterative method,
the PES method is adopted in this paper.

\subsection{Initial and Boundary Conditions}
As an initial condition, we prepare a two-phase atomic gas with a mean density of 
$\nave$ as a precursor of MCs.
To generate it, the following calculation is performed.
A thermally unstable gas with a uniform density of $\nave$ 
is set in a cubic simulation box of 
$(20~\mathrm{pc})^3$ in volume.
We add a velocity perturbation that follows a power law with the Kolmogorov spectral index.
There are two reasons why the Kolmogorov spectrum is used in the initial velocity dispersion.
One is that \citet{Larson1979} observationally discovered that there is a 
correlation between the velocity dispersion $\sigma$ of the atomic gas and the size of regions $l$,
$\sigma = 0.64~\mathrm{km~s^{-1}}\times (l/\mathrm{pc})^{0.37}$. 
The relation is quite similar to the Kolmogorov spectrum, which has the relation 
$\sigma \propto l^{1/3}$.
The other is that the observed turbulence in the atomic gas is subsonic or transonic for the WNM.
Theoretically, it is natural that subsonic turbulence follows the Kolmogorov spectrum, which 
is satisfied in incompressible turbulence.
The spectrum of the initial velocity dispersion relates to the size distribution 
of the clumps of cold neutral medium (CNM), 
which may affect the efficiency of converting the upstream kinetic energy 
into the post-shock kinetic energy. 
The initial velocity dispersion is 10\% of the initial sound speed.
The initial magnetic field ${\bf B}_0$ is set to be uniform. 
The field strength is denoted by $B_0$, and 
the angle between ${\bf B}_0$ and the $x$-axis is denoted by $\theta$.
Imposing periodic boundary conditions in all directions, 
we solve the basic equations until $t=8~$Myr which 
corresponds to several times of the cooling time of the initial unstable gas.
In this calculation, we consider 
the optically thin cooling/heating processes and chemical reactions by
ignoring the dust extinction and optical depth of the cooling photons.
To save the computational time, the calculation is performed with a resolution of $128^3$.

\begin{figure}[h]
     \centering
     \includegraphics[width=9cm]{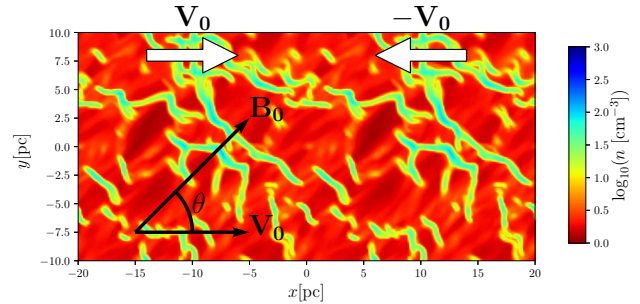}
     \caption{An example of the initial conditions: density cross section 
     at the $z=0$ plane 
     of the initial condition for $\nave=5~$cm$^{-3}$, $B_0=5~\mu$G, and $\theta=45^\circ$} 
     \label{fig:initial}
\end{figure}

Next we set an initial condition for the simulation of 
MC formation using the two-phase atomic gas obtained using the calculation shown above. 
The calculation domain is doubled in the $x$-direction, indicating that the volume 
is $(40~\mathrm{pc})\times(20~\mathrm{pc})^2$.
The physical quantities are copied periodically in the $x$-direction.
Fig. \ref{fig:initial} shows the density slice at the $z=0$ plane of the initial condition for 
$\nave=5~$cm$^{-3}$, $B_0=5~\mu$G, 
$\theta=45^\circ$, and $V_0=20$~km~s$^{-1}$.
A colliding flow profile $v_x(x,y,z)=-V_0\tanh(x/(1~\mathrm{pc}))$ is added in the initial 
profile.
The Dirichlet boundary conditions are imposed at $x=\pm 20~$pc 
in a manner such that the initial distribution is continuously injected into the calculation domain 
with constant velocities of $V_0$ and $-V_0$ from $x=20~\mathrm{pc}$ and $x=-20~\mathrm{pc}$, 
respectively. 
Periodic boundary conditions are imposed at the $y$- and $z$-boundaries. 
The simulations are conducted 
on uniform $1024\times 512\times 512$ cells, leading to a grid size of 0.04~pc.

\subsection{Methods for Estimating Column Densities and Optical Depths}\label{sec:ray}
In order to calculate the FUV flux at an arbitrary point, we need to integrate the 
column density along every ray path 
from the boundaries of the simulation box, summing over the contributions from each path.
Since the exact radiation transfer is computationally expensive, 
a two-ray approximation in the $x$-direction is used \citep{Inoue2012}.
The FUV irradiation is calculated with two rays, which irradiate in the 
$x$-direction from the $x$-boundaries, $x = \pm 20$~pc. The flux of each ray is $G_0=0.5$.
This approximation is justified by the geometry of the MC formation site.
The compressed region has a sheet-like configuration extended in the $(y,z)$ plane. 
FUV photons penetrate mainly in the $x$-direction because the column densities 
in the $x$-direction are
smaller than those along the $y$- and $z$-directions in the sheet structures.

The values of the local visual extinction at $(x,y,z)$ measured from the $x=20$~pc and 
$x=-20$~pc are given by
\begin{equation}
     A_\mathrm{V}^-(x,y,z) = \frac{1}{N_0}\int_{-L}^{x} n(x',y,z) dx'\;\;\;
\end{equation}
and
\begin{equation}
     A_\mathrm{V}^+(x,y,z) = \frac{1}{N_0}\int_{x}^{L} n(x',y,z) dx',
\end{equation}
respectively, where $N_0=1.9\times 10^{21}$~cm$^{-2}$ and $L=20$~pc.

The local column densities of C, H$_2$, and CO used in the chemical reactions and 
heating/cooling processes are calculated 
in a manner similar to that used in calculating
the visual extinction.
The two-ray approximation is also used for 
for the escape probabilities of the [OI] and [CI$\!$I] cooling photons.
The escape probabilities are calculated as follows:
we estimate $\tau^+$ and $\tau^-$, which are the 
optical depths integrated from the left and right boundaries,
respectively,  
\begin{equation}
     \tau^{\pm}(\mathrm{C^+}, \mathrm{O}) = 
     \frac{N^{\pm}(\mathrm{C^+}, \mathrm{O})
     }{N_{\tau}\delta v_5},
\end{equation}
where $N_{\tau} = 2.6\times 10^{17}~\mathrm{cm}^{-2}$ for the [OI] line
and $N_{\tau}=1.5\times 10^{17}~\mathrm{cm}^{-2}$ 
for the [CI$\!$I] line \citep{HM1989}, and
$\delta v_5 = \delta v/(1~\mathrm{km~s^{-1}})$ represents the Doppler shift effect.
The OI and CI$\!$I cooling processes are important in relatively low-density regions 
where the velocity dispersion
along the $x$-direction is as high as $\sim 8$~km~s$^{-1}$. Thus, $\delta v_5 = 8$ is adopted.
The escape probability is evaluated as $\beta(\tau^+) + \beta(\tau^-)$, in which
$\beta(\tau)$ is the escape probability in a semi-infinite medium \citep{deJong1980}.
The optical depth of the CO line cooling is 
evaluated in a similar way,
but $\delta v_5=3$ is used because the high-density regions where 
CO forms have lower velocity dispersions.

\begin{figure*}[htpb]
     \centering
     \includegraphics[width=18cm]{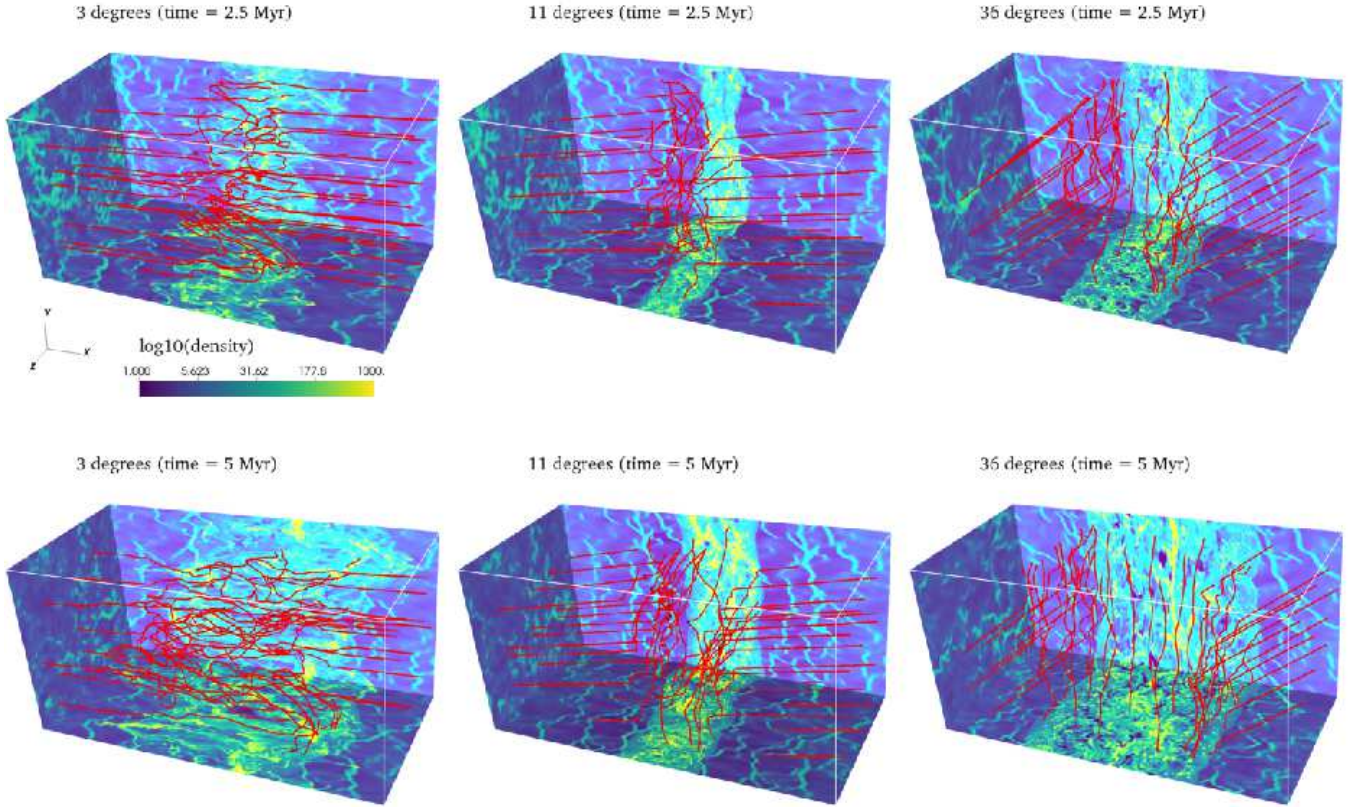}
     \caption{
     Density slices at the three orthogonal planes for $\Theta3$ (left column), $\Theta11$ (middle column), 
     and $\Theta36$ (right column).
     The upper and lower panels correspond to the results at $t=2.5$~Myr and $t=5$~Myr.
     The magnetic field lines are plotted as the red lines.
     }
     \label{fig:3d}
\end{figure*}
\section{Results of a Fiducial Parameter Set}\label{sec:results}
In this section, we investigate how the MC formation depends on $\theta$ in the fiducial 
model with a parameter set of $\nave=5~\pcc$, $V_0=20$~km~s$^{-1}$, $B_0=5~\mu$G.
Super-shells are often observed as HI shells that have expansion speeds of $\sim 10-20~$km~s$^{-1}$ 
and sizes of a few hundred parsecs \citep{Heiles1979}.
Shock compressions with a velocity difference of $2V_0=40$~km~s$^{-1}$ 
are expected in super-shells younger than $10~$Myr \citep{McCray1987}.
The fiducial parameter set is motivated by collisions between adjacent expanding HI shells.
Our setups are also relevant to the 
large-scale converging flows with a speed of a few tens of km~s$^{-1}$ 
associated with the spiral arm formation \citep{Wada2011}.
The field strength $B_0=5~\mu$G is close to the median value $\sim 6~\mu$G measured by
observations of the Zeeman effect in CNM \citep{Heiles2005,Crutcher2010}.
The median field strength is roughly consistent with other observations \citep{Beck2001}.
Note that the median field strength of $\sim 6~\mu$G is not necessarily the most probable one 
since the probability distribution function (PDF) of the field strength is only loosely
constrained by observations in the range $0 < B_0 < 12~\mu$G.
We explore the parameter space of ($\nave$, $V_0$, $B_0$) in Section \ref{sec:parameter}.

The upstream atomic gas has a two-phase structure as shown in Fig. \ref{fig:initial}.
In this paper, we define WNM as the gas with a temperature higher than $10^3$~K.
Typical densities of the CNM clumps and WNM
are $\sim 40~\mathrm{cm}^{-3}$ and $\sim 2~\mathrm{cm}^{-3}$, respectively, 
indicating that the density contrast is as large as $\sim 20$.
The density of the CNM is consistent with observations \citep{Heiles2003b}.
In our initial conditions, 
the CNM clumps make up roughly half of the total mass.
Note that the warm gas, which we call WNM, is not in thermal equilibrium but still 
in a thermally unstable state.
Indeed, observations have revealed that a substantial fraction of the atomic gas is 
in the unstable regime \citep{Heiles2003b,Kanekar2003,Roy2013}.
Theoretically, it is easy for the WNM to 
deviate from the thermal equilibrium state due to turbulence 
because the cooling/heating timescales are long \citep{Gazol2001,AH2005}. 
The Mach numbers of the colliding flow
with respect to the WNM and CNM are $\sim 4$ and $\sim 18$, respectively.
The Alfv{\'e}n Mach number of the colliding flow
for the WNM is $\sim 3$, while that for the CNM is $\sim 14$.

The results of three main models are shown:
one is the case of an almost parallel field of $\theta=3^\circ$ which 
provides almost the same results as in the case of a completely parallel field case 
in \citet{Inoue2012}, and 
the others are cases of oblique field in which the magnetic fields are tilted at 
$\theta=11^\circ$ and $36^\circ$ to the upstream flow.
We name a model by attaching ``$\Theta$'' in front of a value of $\theta$ in degrees, i.e.,
models $\Theta3$, $\Theta11$, and $\Theta36$.

A head-on colliding flow produces two shock fronts that propagate outward. 
The simulations are terminated at $t=5~$Myr because 
the shock fronts reach the $x$-boundaries for model $\Theta3$.
The termination time ($t=5$~Myr) is longer than the cooling times, 
which are approximately 1~Myr for the shocked warm gas and 
$10^{-2}$~Myr for the shocked CNM clumps (Equation (\ref{tc})).
The termination time will be compared with the dynamical time in Section \ref{sec:vdisp}.

\begin{figure}[htpb]
     \centering
     \includegraphics[width=8cm]{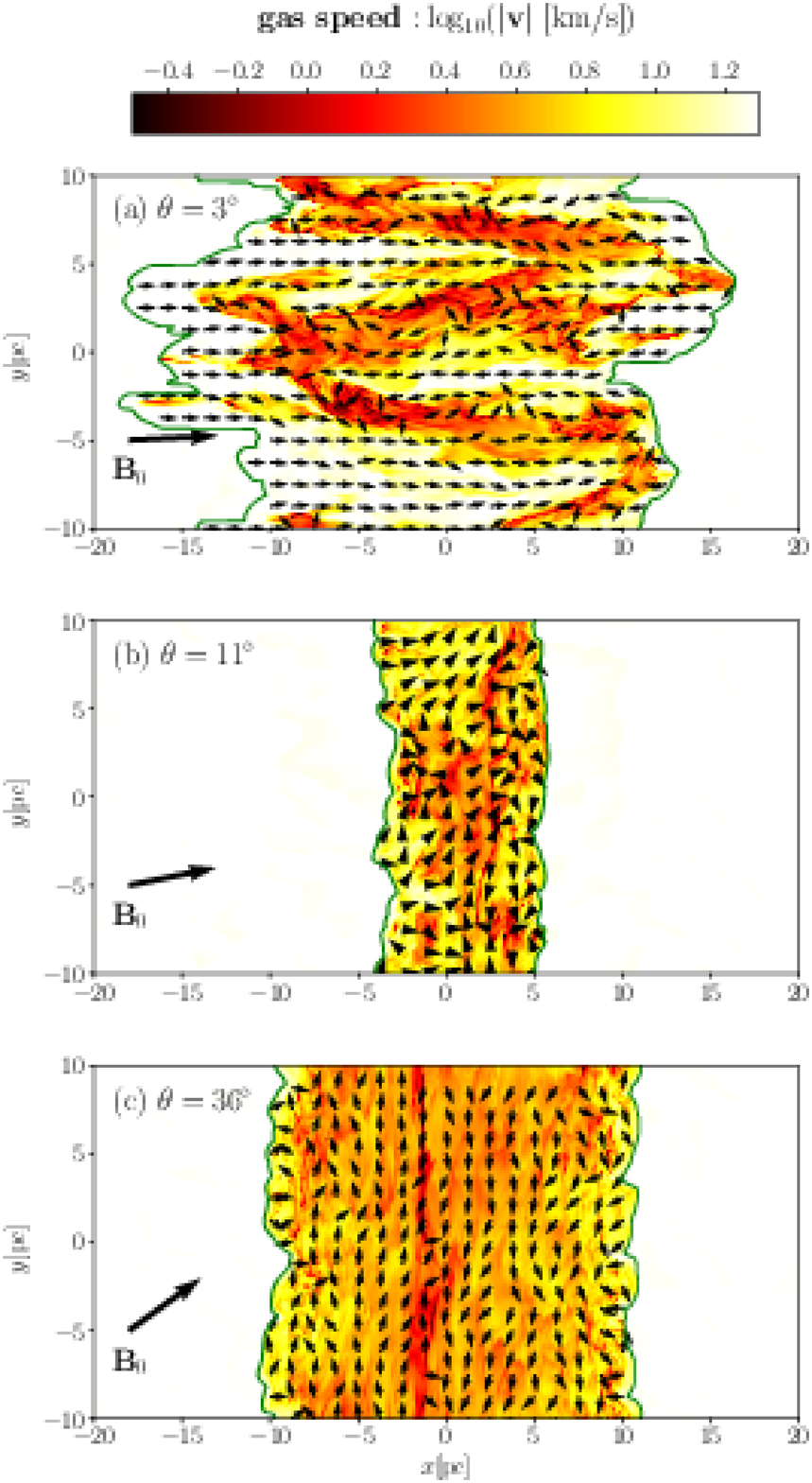}
     \caption{
     Gas speed slices in the $z=0$ plane at $t=5$~Myr for models (a) $\Theta3$, (b) $\Theta11$, and (c) $\Theta36$.
     The velocity field is shown by the arrows only inside the post-shock layer.
The two shock fronts whose positions are defined by the minimum and maximum of 
the $x$ coordinates where 
$P(x,y,z)/k_\mathrm{B}> 5.8\times 10^3~\mathrm{K\cdot cm^{-3}}$ is satisfied.
The positions of the identified shock fronts are shown by the green lines.
     }
     \label{fig:slicez}
\end{figure}
\subsection{Main Features}\label{sec:slice}
Fig. \ref{fig:3d} shows the density color maps 
in the three orthogonal planes at $t=2.5$~Myr and $t=5$~Myr for models $\Theta3$, $\Theta11$, and $\Theta36$.
The magnetic field lines are shown by the red lines.
Unlike in WNM colliding flows, the CNM clumps exist in the upstream gas in the present models.
A CNM clump is not decelerated completely when passing through a shock front, and 
plows the post-shock gas with a large $x$-momentum.
The MHD interaction with the surrounding gas decelerates the CNM clump. 

For model $\Theta3$, since the CNM clumps move almost along the magnetic field,
the Lorentz force does not contribute to their deceleration of the CNM clumps significantly.
A CNM clump that enters into the post-shock layer 
from one of the shock fronts is not decelerated completely, and 
it collides with the shock front on the opposite side.  
Since the upstream CNM clumps accrete onto the post-shock layer from both the $\pm x$-directions,  
the shocked CNM clumps are moving in opposite directions 
along the $x$-axis in the post-shock layer.
This ballistic-like motion of the CNM clumps 
pushes the shock fronts outward and significantly deforms them as shown in 
Fig. \ref{fig:3d}. 
The gas motion significantly widens the post-shock layer.
These behaviors have 
been found by \citet{Inoue2012} and \citet[][without magnetic fields]{Carroll2014}.

Note that if there were no radiative cooling, the CNM clumps would be quickly destroyed and 
mixed with the surrounding warm gas after passing through the shock fronts as shown in \citet{KMC1994}.
In our simulations, 
the radiative cooling, high density contrast, and magnetic fields extend the lifetime 
of the CNM clumps. The CNM clumps even grow through
accretion of the surrounding warm gas due to 
radiative cooling. A detailed discussion is given in Appendix \ref{app:cnm}.

The deformation of the shock fronts allows the kinetic energy of the upstream gas to 
remain almost unchanged after passing through them.
The remaining kinetic energy is available to disturb the deep interior of the post-shock layer.
Fig. \ref{fig:slicez}a shows the gas velocity slice at the $z=0$ plane.
The green lines indicate the two shock fronts whose positions are defined as the minimum and maximum of 
the $x$ coordinates where 
$P(x,y,z)/k_\mathrm{B}\ge P_\mathrm{th}$ is satisfied,
where $P_\mathrm{th}$ is a threshold pressure that should
be larger than the upstream mean pressure $P_0 \sim 4.0\times 10^{3}$~K~cm$^{-3}$. 
A value of $5.8\times 10^3$~K~cm$^{-3}$ is adopted. 
We confirmed that the results are not sensitive to the choice of $P_\mathrm{th}$ as long 
as $P_\mathrm{th}$ is not far from $P_0$.
In this figure, channel-flow-like fast WNM streams are visible in the post-shock layer. 
They are almost aligned to the $x$-axis, and their speeds are as fast as 10~km~s$^{-1}$ which is comparable
to the WNM sound speed. 
The CNM clumps are entrained by interaction with the surrounding warm gas, 
forming a filamentary structure elongated along the $x$-axis
\citep[also see][]{Mellema2002,Cooper2009}.

Magnetic fields are passively bent because the gas motion is super-Alfv{\'e}nic, as will 
be shown in Section \ref{sec:pre}.
The CNM clumps moving in opposite directions stretch the field lines 
preferentially in the direction of collision.
As a result, the field lines are folded (Fig. \ref{fig:3d}).
There are regions where the magnetic fields are flipped 
from the original orientation \citep{Inoue2012}.

The middle column of Fig. \ref{fig:3d} demonstrates that 
the small obliqueness of $\theta=11^\circ$ drastically changes the post-shock structure. 
Since the shock compression amplifies the tangential component of the magnetic field,
the field lines are preferentially aligned to the $y$-axis although 
they have significant fluctuations.
Since the shock-amplified tangential magnetic field pulls the CNM clumps back, 
they are decelerated before reaching the opposite side of the post-shock layer.
The decelerated CNM clumps are accumulated in the central region. 
Unlike for model $\Theta3$, 
there is no significant gas motion across the thickness of the post-shock layer (Fig. \ref{fig:slicez}b). 
The post-shock layer therefore becomes much thinner than that for model $\Theta3$.

When the magnetic field is further tilted at $\theta=36^\circ$ to the upstream flow, 
the results are almost the same as those for model $\Theta11$ but
the post-shock layer is thicker for model $\Theta36$ (Fig. \ref{fig:3d}c).
This is simply because it contains larger magnetic fluxes.

\subsection{Transversely Averaged Momentum Flux 
in the Compression Direction}\label{sec:pre}

Figs. \ref{fig:3d} and \ref{fig:slicez} suggest that 
the magnetic field controls the post-shock structures if 
the angle $\theta$ is large enough.
To investigate the role of the magnetic fields in 
the post-shock structures quantitatively,
we measure the momentum flux along the $x$-axis, which 
consists of the ram pressure 
($P_\mathrm{ram} \equiv \rho v_x^2$), magnetic stress
($\Pi_\mathrm{mag} \equiv (B_y^2 + B_z^2 - B_x^2)/8\pi$), and 
thermal pressure ($P_\mathrm{th}$). 
Note that the magnetic stress $\Pi_\mathrm{mag}$ is not the magnetic pressure but 
the $xx$ component of the Maxwell stress tensor. 
Since the magnetic tension contributes to $\Pi_\mathrm{mag}$,
$\Pi_\mathrm{mag}$ can be negative if the magnetic tension 
is larger than the magnetic pressure.
These quantities are useful for understanding quantitatively which 
pressures are dominant in the post-shock layers.

\begin{figure}[htpb]
     \centering
     \includegraphics[width=8cm]{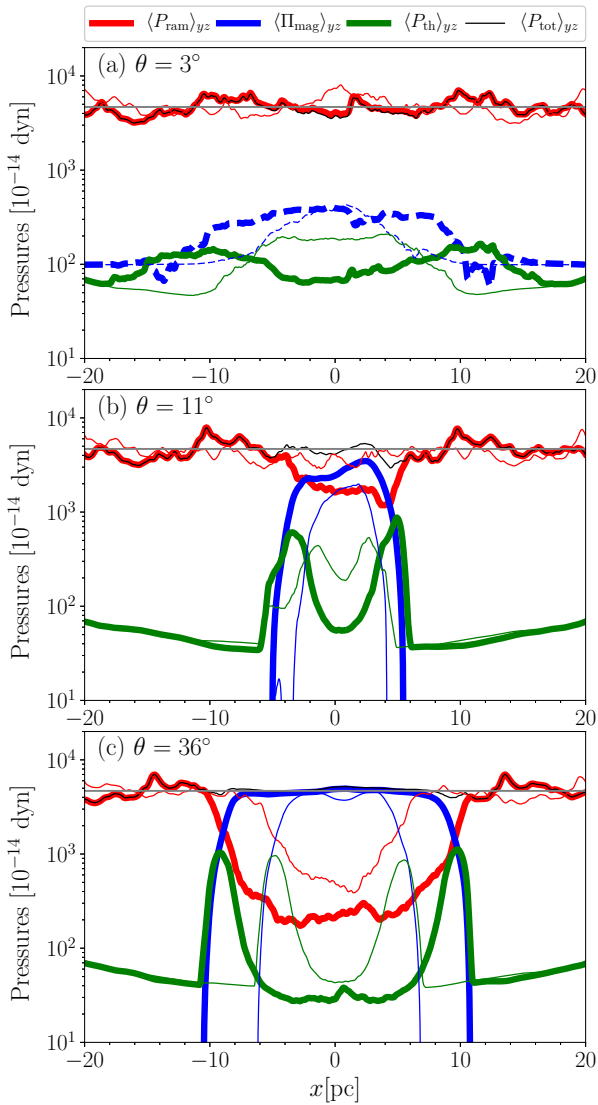}
     \caption{
     Transversely averaged 
     $P_\mathrm{ram}$ (red), 
     $\Pi_\mathrm{mag}$ (blue), and $P_\mathrm{th}$ (green) for 
     models (a) $\Theta3$, (b) $\Theta11$, and (c) $\Theta36$.
     The thin and thick lines represent the results at 
     $t=2.5$~Myr and $t=5$~Myr, 
     respectively.
     The blue dashed line shows that $\Pi_\mathrm{mag}$ is negative, indicating that 
     the magnetic tension is larger than the magnetic pressure.
     The total pressures are shown by the black lines only at $t=5$~Myr.
     The horizontal gray line corresponds to the ram pressure of the atomic gas 
     $\langle \rho_0\rangle V_0^2$.
     }
     \label{fig:1dim_pre}
\end{figure}

The transversely volume-weighted averaged momentum fluxes, 
which are denoted by $\Pturb_{yz}$, $\Pmag_{yz}$, and $\Pth_{yz}$, 
are plotted in Fig. \ref{fig:1dim_pre}
as functions of $x$ at two different epochs, of $t=2.5$~Myr and $t=5$~Myr
for models $\Theta3$, $\Theta11$, and $\Theta36$. 
In all the models, the transversely averaged total momentum flux
($\Ptot_{yz} = \Pturb_{yz} + \Pmag_{yz} + \Pth_{yz}$)
almost coincides with the upstream ram pressure $\langle \rho_0\rangle V_0^2$, 
indicating that the pressure balance is established along the $x$-axis.
This implies that the average propagation speed of the shock fronts is negligible 
in the computation frame.
Indeed, the average shock propagation speed along the $x$-axis is only a few km~s$^{-1}$ 
in the computation frame, and hence the shock velocity with respect to the upstream gas 
does not differ from $V_0$ by more than $\sim 10$\%.

For model $\Theta3$, 
$\Pturb_{yz}$ is much larger than the other pressures throughout the post-shock layer, regardless of time.
This clearly shows that the gas motion is highly supersonic and super-Alfv\'enic. 
The transversely averaged magnetic stress $\Pmag_{yz}$ is always negative, or 
$\langle B_x^2 \rangle > \langle B_y^2+B_z^2\rangle$ because
the parallel component of the magnetic field 
is preferentially amplified by the channel-like fast gas flows biased in the collision direction 
(Fig. \ref{fig:slicez}a). 
Amplification of the transverse field component does not work significantly.

For the oblique field models ($\Theta11$ and $\Theta36$),
the post-shock layers can be divided into two regions.
One is the warm surface layer, where all the pressures 
contribute equally to the total momentum flux.
The other is the cold central layer, where $\Pth_{yz}$ is low 
while $\Pmag_{yz}$ is high.
A similar two-region structure was also found in WNM colliding flows perpendicular to 
the magnetic field \citep{Heitsch2009}.

The warm surface layers are formed by a continuous supply of the upstream WNM.
After shock heating, it cools down by radiative cooling, and accretes onto the cold central layer.
Thus, the thicknesses of the warm surface layers are roughly determined by the cooling length;
this is estimated to be 2~pc, which is comparable to their thicknesses in Fig. \ref{fig:1dim_pre}.
When a CNM clump enters the post-shock layer, it passes through a warm surface layer easily and 
collides with the cold central layer.

In the cold central layer, $\Pturb_{yz}$ decreases with time both for models $\Theta11$ and $\Theta36$.
This is because the cold central layer is ``shielded'' by the tangential magnetic field 
shown in Fig. \ref{fig:3d} against the accretion of the CNM clumps.
This accretion of the CNM clumps disturbs the gas  near the surfaces of the cold central layer.
The turbulence inside the cold central layer decays due to numerical dissipation 
and escaping cooling photons from the shock-heated regions, 
leading to a decrease in $\Pturb_{yz}$ in the central layer.
The time evolution of $\Pmag_{yz}$ in the central layer is different between models $\Theta11$ and $\Theta36$.
For model $\Theta11$, at $t=2.5$~Myr, $\Pturb_{yz}$ plays an important role in the total pressure. 
Thus, the decrease in $\Pturb_{yz}$ causes 
the post-shock layer to become denser, leading to an
increase in $\Pmag_{yz}$ due to flux freezing (Fig. \ref{fig:1dim_pre}b).
By contrast, 
for model $\Theta36$, the post-shock layer is
mainly supported by the magnetic stress at $t=2.5$~Myr.
Since $\Pturb_{yz}$ makes a negligible 
contribution to the pressure balance where $\Pmag_{yz} \sim \rhoave V_0^2$,
$\Pmag_{yz}$ is constant 
with time in the cold central layer (Fig. \ref{fig:1dim_pre}c).

\subsection{Velocity Dispersion}\label{sec:vdisp}
In Section \ref{sec:pre}, we found that the ram pressure does not decrease for model $\Theta3$  
while it decreases with time for the oblique field models ($\Theta11$ and $\Theta36$).
In this section, we examine the time evolution of the post-shock velocity dispersion.

\begin{figure}[htpb]
     \centering
     \includegraphics[width=8cm]{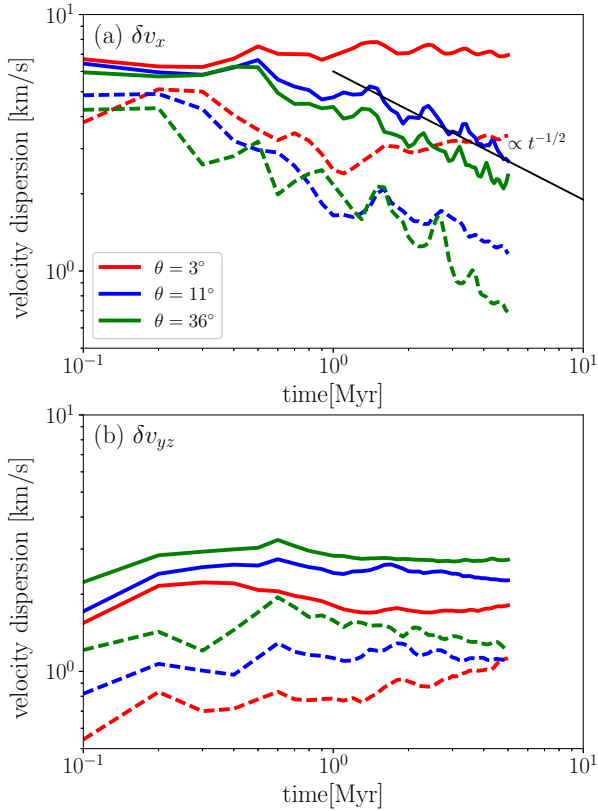}
     \caption{
             Time evolution of the mass-weighted (solid lines) 
             and CO-density-weighted (dashed lines) velocity dispersions 
     for models $\Theta3$ (red), $\Theta11$ (blue), and $\Theta36$ (green).
     The top and bottom panels represent the velocity dispersions parallel to the $x$-axis and 
     those averaged in the $y$- and $z$-components, 
     $\delta v_{yz} \equiv  \sqrt{(\delta v_y^2 + \delta v_z^2)/2}$, respectively.
     Equation (\ref{vx}) is shown by the black line in the top panel. 
     }
     \label{fig:vdisp}
\end{figure}

Fig. \ref{fig:vdisp} shows the mass-weighted and CO-density-weighted velocity dispersions,
which are measured along the $x$-axis ($\delta v_x$) 
and averaged between the $y$- and $z$-components ($\delta v_{yz} \equiv\sqrt{(\delta v_y^2 + \delta v_z^2)/2}$) 
for models $\Theta3$, $\Theta11$, and $\Theta36$.
The CO-density-weighted velocity dispersions correspond to the velocity dispersions in the dense regions 
where $n>10^3~\pcc$ because CO forms only in the dense regions as shown in Section \ref{sec:mol}. 

We investigate how many dynamical timescales are considered in the simulations.
The dynamical time depends on the direction.
In the $x$-direction, the dynamical time is given by $t_{\mathrm{dyn},x} = L_x/\delta v_x$, 
where $L_x$ is the width of the post-shock layer.
By measuring $L_x$ from Fig. \ref{fig:slicez} and $\delta v_x$ from Fig. \ref{fig:vdisp}a 
at $t=5$~Myr, one obtains 
$t_{\mathrm{dyn},x} \sim 3~$Myr for models $\Theta3$ and $\Theta11$, and 
$6.5$~Myr for model $\Theta36$.
Thus, the post-shock gases are mixed well on the $x$-axis for models $\Theta3$ and $\Theta11$. 
The dynamical times with respect to the transverse direction are
given by $({20~\mathrm{pc}})/\delta v_{yz} \sim 8~\mathrm{Myr}$ where $\delta v_{yz}$ is set to 
2.5~km~s$^{-1}$ from Fig. \ref{fig:vdisp}b for all the models. 
The simulations are terminated before the transverse dynamical time.

\begin{figure}[htpb]
     \centering
     \includegraphics[width=8cm]{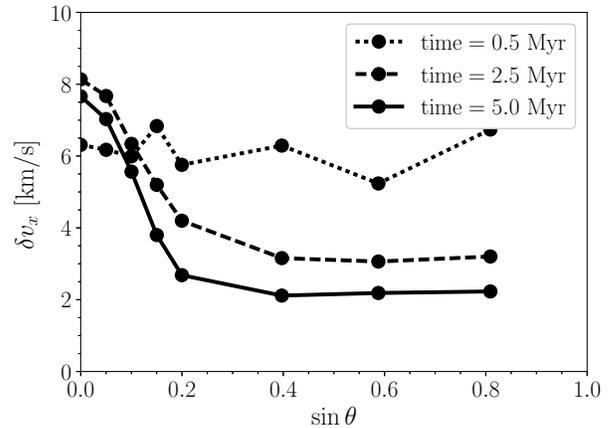}
\caption{
    The velocity dispersions parallel to the $x$-axis measured 
    at the three different epochs $t=0.5~$Myr (dotted), 
    $t=2.5$~Myr (dashed), and $t=5~$Myr (solid) are plotted as a function of $\sin \theta$.
    }
     \label{fig:dvx_th}
\end{figure}

\begin{figure}[htpb]
     \centering
     \includegraphics[width=8cm]{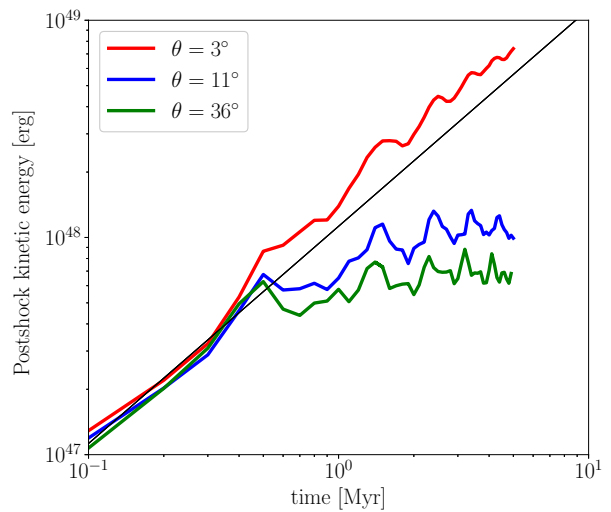}
     \caption{
     Time evolution of the total post-shock kinetic energies parallel to the $x$-axis
     for models $\Theta3$ (red), $\Theta11$ (blue), and $\Theta36$ (green).
     The black line indicates the time evolution of the post-shock kinetic energy 
     in the case where the fraction $\epsilon =10~\%$ of the upstream kinetic energy is 
     converted into the post-shock kinetic energy
     $(\epsilon \dot{M}_\mathrm{tot}V_0^2/2)t$.
     }
     \label{fig:kinene}
\end{figure}

First, we examine the velocity dispersion parallel to the $x$-axis.
Fig. \ref{fig:vdisp}a shows that in the early phase ($t<0.5~$Myr), 
$\delta v_x$ is as high as $6-7~$km~s$^{-1}$.
Fig. \ref{fig:dvx_th} indicates that 
$\delta v_x(t=0.5~\mathrm{Myr})$ is almost independent of $\theta$ although 
there are fluctuations.
The efficiency $\epsilon$ for converting the kinetic energy of the upstream atomic gas 
into the post-shock kinetic energy parallel to the collision direction
is expressed as
\begin{equation}
\epsilon \equiv \frac{\dot{M}_\mathrm{tot}t\delta v_x^2/2}
{\dot{M}_\mathrm{tot}t  V_0^2/2}
     = \left(\frac{\delta v_x}{V_0}\right)^2 \sim 10\%,
     \label{eps}
\end{equation}
where $\dot{M}_\mathrm{tot} = 2\rhoave V_0 L^2$ is the mean mass accretion rate.
Fig. \ref{fig:kinene} illustrates that the time evolution of the total 
post-shock kinetic energies parallel to the $x$-axis, which increase obeying
$(\epsilon \dot{M}_\mathrm{tot} V_0^2/2) t$ for all the models in the early phase, where $\epsilon=10\%$.
The efficiency $\epsilon=10\%$
is larger than those obtained from simulations considering WNM colliding flows 
\citep{Heitsch2009,Kortgen2015,Zam2018}. 
The upstream two-phase structure enhances the efficiency.

For $t>0.5$~Myr, we find a clear dependence of $\delta v_x$ on $\theta$ from Fig. \ref{fig:vdisp}a.
The velocity dispersion parallel to the $x$-axis is almost constant 
or even slightly increases with time 
at least until $5$~Myr for model $\Theta3$,
while it decreases with time for models $\Theta11$ and $\Theta36$.

Why does $\delta v_x$ not decrease for model $\Theta3$?
One reason is that the gas flows are not fully turbulent, but 
rather laminar as shown in Fig. \ref{fig:slicez}a.
The gas flows are strongly biased in the collision direction, and the transverse motion is 
restricted ($\delta v_{yz}/\delta v_{x}\sim 0.3$ in Figs. \ref{fig:vdisp}a and \ref{fig:vdisp}b).
Since translational coherent gas flow is unlikely to decay, 
$\delta v_x$ does not decrease.

Interestingly, for both of the models $\Theta11$ and $\Theta36$, 
the velocity dispersions parallel to the $x$-axis are closely approximated by 
\begin{equation}
\delta v_x \sim \delta v_{x0} (t/t_0)^{-1/2},
\label{vx}
\end{equation}
where $\delta v_{x0}=6~\mathrm{km~s^{-1}}$ and $t_0=1~\mathrm{Myr}$ although 
$\delta v_x$ is slightly smaller for model $\Theta36$ than for model $\Theta11$.
In order to see the universality of the time evolution of $\delta v_x$ more clearly, 
we investigate the $\theta$-dependence of $\delta v_x$ using
additional simulations with different field orientations.
Fig. \ref{fig:dvx_th} shows $\delta v_x$ at the three different epochs as a function of $\sin \theta$.
At both $t=2.5$~Myr and at $t=5~$Myr, $\delta v_x$ shows a decrease in the range of 
$0^\circ<\theta<11^\circ$, and the decreases suddenly slow down around $\theta\sim 11^\circ$. 
For angles larger than $\theta\sim 11^\circ$, $\delta v_x$ does not depend on $\theta$ sensitively. 

The time dependence of $\delta v_x\propto t^{-1/2}$ implies 
that the total kinetic energy parallel to the $x$-axis 
($\dot{M}_\mathrm{tot}t\delta v_x^2/2$) is constant with time.
Fig. \ref{fig:kinene} indicates  
that the total post-shock kinetic energies parallel to the $x$-axis 
are constant with time in the late phase ($t>0.5~$Myr) for models $\Theta11$ and $\Theta36$.
This is probably explained if  
the rate of kinetic energy input from the shock fronts 
balances with dissipation rates.

The transverse velocity dispersion $\delta v_{yz}$ shows the opposite trend to $\delta v_x$;
$\delta v_{yz}$ is larger for larger $\theta$ although the difference is small.
The reason why model $\Theta3$ gives the lowest $\delta v_{yz}$ is that 
the motion of the CNM clumps is not randomized in the post-shock layer (Fig. \ref{fig:slicez}a).
In contrast to $\delta v_x$, $\delta v_{yz}$ does not decrease 
even for models $\Theta11$ and $\Theta36$.
There are several mechanisms to drive transverse velocity dispersion. 
Especially for larger angles (e.g., model $\Theta36$),
the presence of an upstream 
field component perpendicular to the collision direction 
drives a transverse flow behind the shock front following 
the MHD Rankine-Hugoniot relation \citep{deHoffmann1950}.
The transverse flow is generated even without any perturbation.
In Fig. \ref{fig:slicez}c, 
the gas is moving coherently in the $+y$-direction for $x<0$ and in the $-y$-direction for $x>0$.
As long as a colliding flow is stationary, the transverse flow speed
should remain approximately constant. 
Another mechanism is that the shock-amplified magnetic field bends the gas motion so that 
the gas flow is parallel to the transverse direction \citep{Heitsch2009}.
The shock deformation due to the accretion of the CNM clumps 
generates a transverse flow along the magnetic field \citep{Inoue2013,Inoue2018}.
The thermal instability that develops preferentially along the magnetic field 
\citep{Field1965} also contributes to the transverse velocity dispersion.


For all the models, the CO-density-weighted velocity dispersions are larger than $1$~km~s$^{-1}$, 
indicating that the velocity dispersions in the cold dense gases are supersonic 
with respect to their sound speeds ($\sim 0.3~$km~s$^{-1}$).
Collision of the two-phase atomic gas drives stronger post-shock turbulence than 
that of the WNM \citep{Inoue2012,Carroll2014}.

\subsection{Mean Post-shock Densities}\label{sec:meanden}
\begin{figure}[htpb]
     \centering
     \includegraphics[width=8cm]{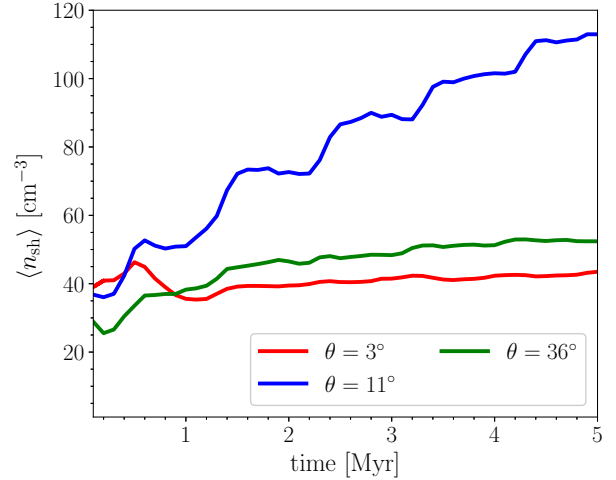}
     \caption{
     Time evolution of the mean post-shock densities
     for models $\Theta3$ (red), $\Theta11$ (blue), and $\Theta36$ (green).
     }
     \label{fig:timeden}
\end{figure}

Fig. \ref{fig:timeden} shows the time evolution of the mean post-shock densities 
$\nsh$ which are derived by averaging densities over the identified post-shock layers.

In the early phase ($t<0.5~\mathrm{Myr}$), 
the mean post-shock densities are enhanced only by a factor of 8 and 
are comparable among the models 
because the super-Alfv{\'e}nic collision of 
the inhomogeneous atomic gas drives strong velocity dispersions, regardless 
of $\theta$ (Fig. \ref{fig:vdisp}a). 

Fig. \ref{fig:timeden} shows that 
only model $\Theta11$ exhibits a rapid increase in $\nsh$ while the mean post-shock densities retain
their initial values for the other models. 
For model $\Theta11$, the obliqueness is large enough for the cold central layer to develop, but 
it is small enough for $\Pturb$ to be a main contributor to the support of 
the post-shock layer against the upstream ram pressure.
Thus, the decrease in $\Pturb$ causes the post-shock layer to 
become denser, leading to the increase in $\nsh$.
For model $\Theta3$ ($\Theta36$), the ram pressure (magnetic stress) 
suppresses further gas compression.

\subsection{Density PDFs and Dense Gas Fractions}\label{sec:densh}

We found that the mean post-shock densities increase only for model $\Theta11$ 
from Fig. \ref{fig:timeden}.
In the MC formation, how much dense gas forms is important because molecules are preferentially formed 
in dense gases.
In this section, we investigate the density PDFs and 
the time evolution of the mass fraction of dense gases.

\subsubsection{Density PDFs}
\begin{figure}[htpb]
     \centering
     \includegraphics[width=8cm]{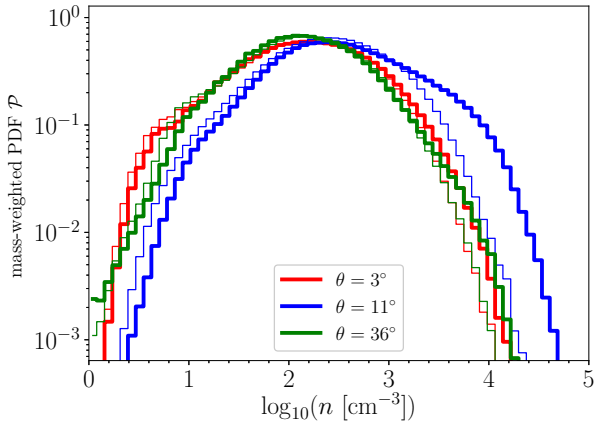}
     \caption{
     Mass-weighted probability distribution functions (PDFs) of gas density
     for models $\Theta3$ (red), $\Theta11$ (blue), and $\Theta36$ (green).
     For each of the models, the thin and thick lines correspond to 
     the results at $t=2.5$~Myr and $t=5$~Myr, respectively.
     }
     \label{fig:denhist}
\end{figure}

The mass-weighted density probability distribution functions (PDFs) 
${\cal P}$ are shown 
in Fig. \ref{fig:denhist} for models $\Theta3$, $\Theta11$, and $\Theta36$.
The definition of ${\cal P}$ is given by 
\begin{equation}
        {\cal P}(X_i) = \frac{1}{M_\mathrm{tot}\Delta X}\int_{ |X-X_i|<\Delta X/2} \rho d^3x,
\end{equation}
where $X\equiv\log_{10}n$ is divided into equally spaced bins whose widths are 
denoted by $\Delta X$, 
$i$ is the index of the bins, and $M_\mathrm{tot}=\dot{M}_\mathrm{tot}t$ 
is the total mass of the post-shock layer.
At $t=2.5$~Myr, the PDFs for all the models have log-normal shapes, although 
the PDF for model $\Theta11$ is slightly shifted toward higher densities 
than for the other PDFs. 
Therefore, model $\Theta11$ has the highest mean density 
(Fig. \ref{fig:timeden}).

For models $\Theta3$ and $\Theta36$, 
the PDFs show little time variation,
as in the time evolution of $\nsh$ found in Fig. \ref{fig:timeden},
although the high-density tails are slightly extended toward higher densities.

By contrast, 
model $\Theta11$ shows significant time variation not only in $\nsh$
but also in the PDF.
Interestingly, even without self-gravity, Fig. \ref{fig:denhist} shows that 
the mass fraction of the dense gas with $n>10^3~\pcc$ significantly increases with time.
This implies that the gas is compressed not only in the collision direction but also 
along the field lines.
The increase in $f_{>10^3}$ is caused by 
transverse flows generated behind a shock front and
by gas condensation due to the thermal instability that tends to develop 
along magnetic fields \citep{Field1965}.

\subsubsection{Time Evolution of Dense Gas Mass Fractions}
To investigate the evolution of dense gases more clearly, 
we measure the mass fractions of the dense gases
with $n>10^2~\pcc$ and $n>10^3~\pcc$, which are 
denoted by $f_{>10^2}$ and $f_{>10^3}$, respectively.

\begin{figure}[htpb]
     \centering
     \includegraphics[width=9cm]{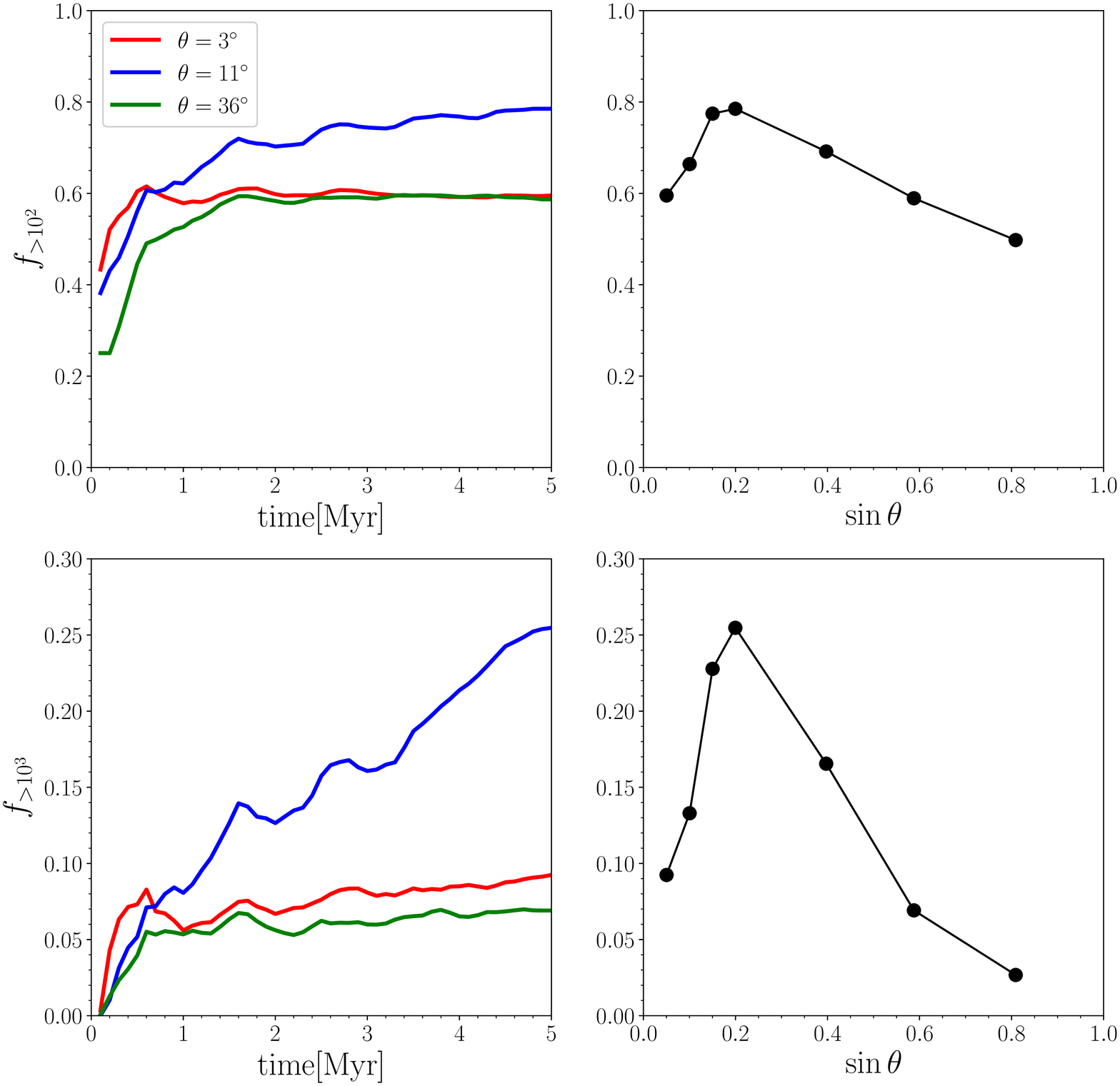}
     \caption{
         Time evolution of the mass fraction of dense gases with $n>10^2~\pcc$ {(the top left panel)}
         and with $n>10^3~\pcc$ {(the bottom left panel)}.
         In both panels, the red, blue, and lines correspond to the results 
         for models $\Theta3$, $\Theta11$, and $\Theta36$, respectively.
         The mass fraction of dense gases with $n>10^2~\pcc$ {(the top right panel)} 
         and $n>10^3~\pcc${(the bottom right panel)} at $t=5~$Myr 
         as a function of $\sin\theta$.
     }
     \label{fig:mass}
\end{figure}

The top left panel of Fig. \ref{fig:mass} shows 
the time evolution of $f_{>10^2}$ for models $\Theta3$, $\Theta11$, and $\Theta36$.
In all the models $f_{>10^2}$ does not depend on time sensitively. 
The dense gas mass with $n>10^2~\pcc$ increases at a constant rate.

Note that the fractional difference in $f_{>10^2}$ is only $30\%$ among the models,
although $\nsh$ for model $\Theta11$ is more than twice as large as 
than those for models $\Theta3$ and $\Theta36$ (Fig. \ref{fig:timeden}).
This indicates that the post-shock layers for models $\Theta3$ and $\Theta36$ contain
wider and less-dense regions than model $\Theta11$ while 
the total masses with $n>10^2~\pcc$ are comparable.

Using the results with various field orientations shown in Fig. \ref{fig:dvx_th},
we plot $f_{>10^2}$ at $t=5$~Myr 
as a function of $\sin\theta$ in the top right panel in Fig. \ref{fig:mass}.
The figure exhibits a weak $\theta$-dependence of $f_{>10^2}$.
The mass fraction has a broad peak around $\theta\sim 11^\circ~(\sin\theta\sim 0.2)$ which 
gives the maximum mean density in Fig. \ref{fig:timeden}.

By contrast, 
the mass fraction of denser gas ($n>10^3~\pcc$) is sensitive to $\theta$.
The time evolution of the mass fraction of the dense gas with $n>10^3~\pcc$ 
is plotted in the bottom left panel of Fig. \ref{fig:mass}.
While $f_{>10^3}$ is constant with time for models $\Theta3$ and $\Theta36$,
$f_{>10^3}$ increases rapidly with time for model $\Theta11$. 
This rapid increase 
is related to the development of the high-density tail in the density PDF for model $\Theta11$ 
(Fig. \ref{fig:denhist}).

The bottom right panel of Fig. \ref{fig:mass} shows that 
$f_{>10^3}$ at $5~\mathrm{Myr}$ has a sharp peak around $\theta\sim 10^\circ$ $(\sin\theta\sim0.2)$. 
For angles larger (smaller) than $\sim 10^\circ$, $f_{>10^3}$ rapidly decreases because 
of the magnetic stress (ram pressure).

\subsection{The Formation of Molecules}\label{sec:mol}
At $t=5$~Myr, the mean accumulated column density reaches $2\nave V_0 t\sim 3.2\times 10^{21}~\mathrm{cm}^{-2}$,
corresponding to a mean visual extinction of $2\nave V_0 t/N_0 \sim  1.6$.
Thus, H$_2$ and CO are expected to form in regions where FUV photons are shielded.
For H$_2$, the self-shielding is effective if column densities of 
H$_2$ are larger than $\sim 10^{14}~$$\mathrm{cm}^{-2}$ \citep{Draine1996}. 
Thus almost all the regions in the post-shock layers are self-shielded.
By contrast, the formation of CO proceeds only when $A_\mathrm{V}$ exceeds unity,
because it requires dust extinction.

\begin{figure}[htpb]
     \centering
     \includegraphics[width=9cm]{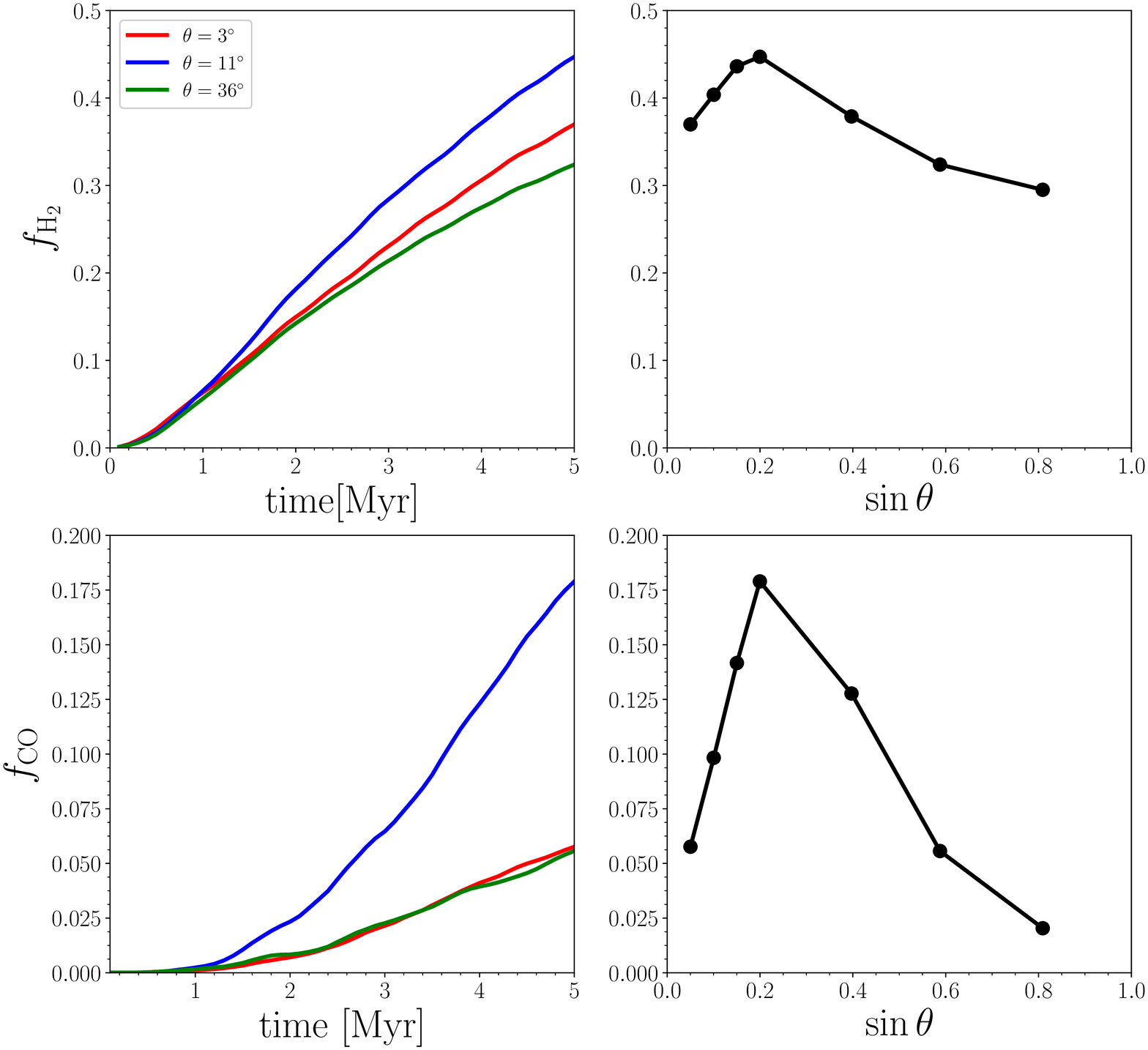}
     \caption{
Time evolution of the H$_2$ fraction {(the top left panel)}
and CO fraction {(the bottom left panel)} in the post-shock layers.
In both panels, the red, blue, and green lines correspond to the results 
for models $\Theta3$, $\Theta11$, and $\Theta36$, respectively.
The H$_2$ fraction {(the top right panel)} and 
CO fraction {(the bottom right panel)} at $t=5~$Myr are plotted as functions of $\sin\theta$.
     }
     \label{fig:H2}
\end{figure}
\begin{figure}[htpb]
     \centering
     \includegraphics[width=8cm]{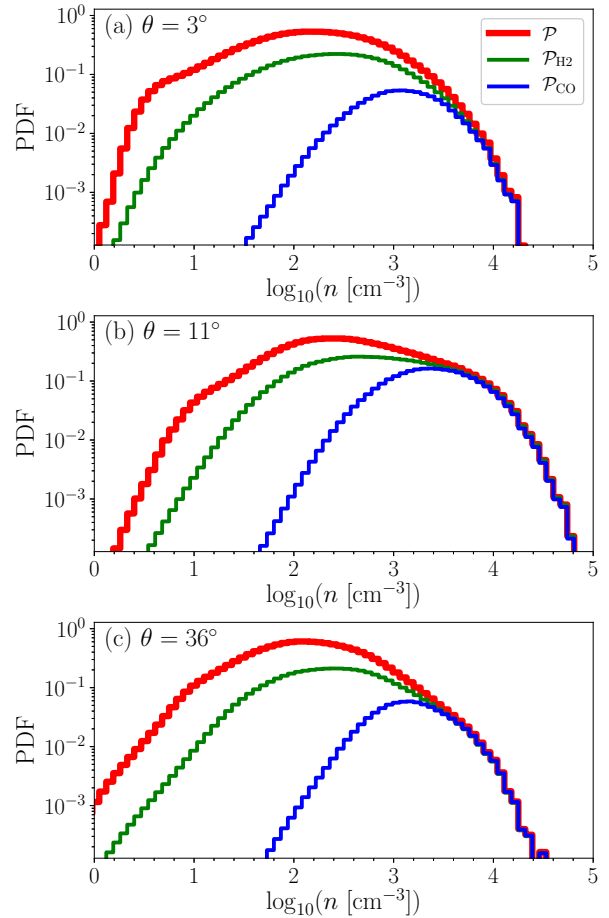}
     \caption{
             Mass-weighted 
             probability distribution function of gas density 
             {(red lines)} for the models (a) $\Theta3$, (b) $\Theta11$, 
             and (c) $\Theta36$.
             The green and blue lines indicate 
   	       the H$_2$-density-weighted (${\cal P}_\mathrm{H_2}$) and 
             CO-density-weighted (${\cal P}_\mathrm{CO}$) PDFs of gas density, respectively.
             In all the panels, the PDFs are calculated at $t=5$~Myr.
     }
     \label{fig:h2co}
\end{figure}
\subsubsection{Formation of hydrogen molecules}
The top left panel of Fig. \ref{fig:H2} shows 
the time evolution of the mass fraction of H$_2$ in the accreted hydrogen 
nuclei, which is defined as
\begin{equation}
     f_\mathrm{H_2} \equiv 
     \frac{\displaystyle \int 2n(\mathrm{H_2}) d^3x}{2\langle n_0 \rangle V_0 L^2 t}.
\end{equation}
The mass fraction of H$_2$ increases continuously with time, and 
the increase in $f_\mathrm{H_2}$ is faster for model $\Theta11$ than 
for models $\Theta3$ and $\Theta36$.
Although the self-shielding is effective, 
the H$_2$ formation on dust grains takes a relatively long time.
A typical formation time of H$_2$ in the gas with a density of $n$ is 
estimated from $t_\mathrm{form,H_2} = (k_\mathrm{H_2}n)^{-1} 
\sim 15~\mathrm{Myr} \left( n/10^2~\mathrm{cm}^{-3} \right)^{-1}$, where 
$k_\mathrm{H_2} \sim 2\times 10^{-17}~\mathrm{cm^3~s^{-1}}$ 
is the H$_2$ formation rate assuming that 
the gas and dust temperatures are 100~K and 10~K, respectively \citep{HM1979}.
\citet{Glover2007} and \citet{Valdivia2016} reported that 
the density inhomogeneity promotes the H$_2$ formation because
$t_\mathrm{H_2}$ is shorter for denser gases.
In order to investigate the effect of the density inhomogeneity 
on the H$_2$ formation, 
we estimate an H$_2$ fraction by assuming that 
the post-shock layer has a spatially and temporally constant density of $n$.
In the derivation of the H$_2$ fraction, we take into account the effect of 
mass accretion, which continuously supplies H$_2$-free atomic gas into 
the post-shock layer.
The detailed derivation is presented in Appendix \ref{app:H2}.
We obtain the H$_2$ fraction at $t_\mathrm{f}=5~$Myr which is given by 
\begin{equation}
        f_\mathrm{H_2,ave}(n) = 1-
    \frac{1 - e^{-2k_\mathrm{H_2} n t_\mathrm{f}}}
   {2k_\mathrm{H_2}n t_\mathrm{f}}.
   \label{fH2}
\end{equation}
For model $\Theta11$, 
$\nsh$ increases from $\sim 40~\pcc$ to $\sim 115~\pcc$ (Fig. \ref{fig:timeden}).
Thus, the H$_2$ fraction derived assuming the spatially constant post-shock density of $\nsh$ 
is expected to take 
a value between $f_\mathrm{H_2,ave}(40~\pcc)=0.1$ and $f_\mathrm{H_2,ave}(115~\pcc)=0.3$.
For models $\Theta3$ and $\Theta36$, $\nsh$ remains a value of 
$\sim 40~\pcc$ from Fig. \ref{fig:timeden}, leading 
to $f_\mathrm{H_2,ave}(40~\pcc)\sim 0.1$. 
Thus for all the models, the formation of H$_2$ proceeds 
faster than predicted by $f_\mathrm{H_2,ave}(n=\nsh)$.


The comparison of $f_\mathrm{H_2}(t=t_\mathrm{f})$ with $f_\mathrm{H_2,ave}$ suggests 
that the rapid formation of H$_2$ arises from the 
high density-inhomogeneity in the post-shock layers.
In order to examine which density ranges are responsible for the H$_2$ formation,
we plot the H$_2$-density-weighted PDFs $P_\mathrm{H_2}$ calculated at $t=5$~Myr for 
the three models in Fig. \ref{fig:h2co}. 
$P_\mathrm{H_2}$ is defined by 
\begin{equation}
{\cal P}_\mathrm{H_2}(X_i) = \frac{\displaystyle
\int_{ |X-X_i|<\Delta X/2} 2 \mu n(\mathrm{H_2})
m_\mathrm{H} d^3x
}{M_\mathrm{tot}\Delta X}.
\end{equation}
For reference, the mass-weighted PDFs of gas density ${\cal P}$ are plotted in Fig. \ref{fig:h2co}.
If ${\cal P}_\mathrm{H_2}$ coincides with ${\cal P}$ at a density bin,
all the hydrogen nuclei become H$_2$ in the corresponding density range.
Fig. \ref{fig:h2co} illustrates that the formation of H$_2$ depends on density.
The H$_2$ fraction is almost unity in the gas with a density larger than $10^3~\pcc$.
This is because $t_\mathrm{H_2}(n>10^3~\pcc$) is less than $1.5$~Myr which is 
sufficiently short to form H$_2$.
The density giving $t_\mathrm{form,H_2} = 5~$Myr is approximately
$300~\pcc$ which 
corresponds to the peak densities of ${\cal P}_\mathrm{H_2}$ 
for all the panels of Fig. \ref{fig:h2co}.


The top right panel of Fig. \ref{fig:H2} shows the $\theta$-dependence of $f_\mathrm{H_2}$ at $t=5$~Myr.
The mass fraction of $\mathrm{H_2}$ does not depend on $\theta$ sensitively, 
and there is
a broad peak around $\theta=11^\circ$.
Even in models $\Theta3$ and $\Theta36$ having low $\nsh$, 
the H$_2$ fractions are comparable to that in model $\Theta11$.
The weak $\theta$-dependence of $f_\mathrm{H_2}$ 
comes from the fact that the H$_2$ formation occurs mainly in 
the dense gas with $n>10^2~\pcc$ whose mass fraction exhibits a weak $\theta$-dependence  
as shown in Fig. \ref{fig:mass}.

\subsubsection{Formation of CO}

Fig. \ref{fig:h2co} shows the CO-density-weighted PDFs of gas density ${\cal P}_\mathrm{CO}$ which 
is defined as
\begin{equation}
       {\cal P}_\mathrm{CO}(X_i) = \frac{\displaystyle
        \int_{ |X-X_i|<\Delta X/2} \mu n(\mathrm{CO}) m_\mathrm{H} d^3x
       }{ {\cal A}_\mathrm{C}M_\mathrm{tot}\Delta X}.
\end{equation}
If ${\cal P}_\mathrm{CO}$ coincides with ${\cal P}$ in a bin,
all the carbon nuclei are in the form of CO in the corresponding density range.
Fig. \ref{fig:h2co} shows that CO formation proceeds preferentially in denser gases  
than H$_2$ formation. For $n<10^3~\pcc$, the CO fractions quickly decrease as density decreases
for all the models
because CO molecules are destroyed by 
FUV photons for low-density regions where $A_\mathrm{V}<1$.
\begin{figure}[htpb]
     \centering
     \includegraphics[width=8cm]{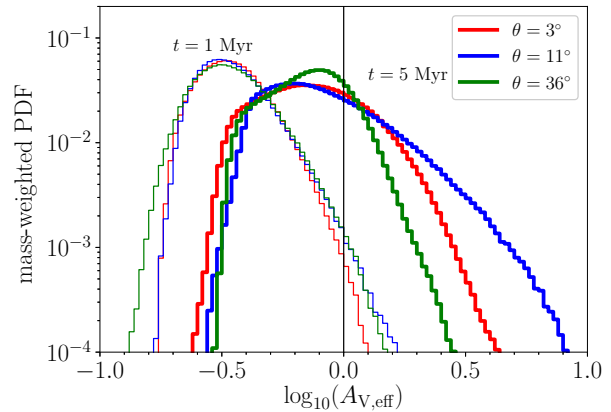}
     \caption{
             Mass-weighted probability distribution function of the effective visual extinction
             for the models $\Theta3$ (red), $\Theta11$ (blue), and $\Theta36$ (green).
             The results at $t=1$~Myr and $t=5$~Myr are shown by the thin and thick lines for 
             each model, respectively.
             The vertical line indicates $A_\mathrm{V,eff}=1$ above which dust extinction works effectively.
     }
     \label{fig:av}
\end{figure}

The bottom left panel of Fig. \ref{fig:H2} shows 
the time evolution of the mass fraction of CO in the accreted C-bearing species, which is defined as
\begin{equation}
     f_\mathrm{CO} = \frac{\int n(\mathrm{CO})  d^3x}{2\nave {\cal A}_\mathrm{C} V_0 L^2 t}.
\end{equation}
The CO fractions remain extremely small in the early phase and 
begin to increase at points near $t\sim 1~$Myr for all the models (the bottom left panel of Fig. \ref{fig:H2}).
This behavior of $f_\mathrm{CO}$ can be roughly understood from the time evolution of dust extinction.
In order to characterize the dust extinction at each position, 
we define the effective visual extinction $A_\mathrm{V,eff}$ as
\begin{equation}
        A_\mathrm{V,eff} = - \frac{1}{2.5} \ln \left[ \frac{1}{2} 
        \left( e^{-2.5A_\mathrm{V}^-} + e^{-2.5A_\mathrm{V}^+} \right) \right],
        \label{av}
\end{equation}
where the factor of $2.5$ comes from the dust extinction factor $e^{-2.5A_\mathrm{V}}$ of the 
CO photodissociation rate \citep{Glover2010}.
In the initial conditions, the mean values of $A_\mathrm{V,eff}$ in the simulation box 
are 0.16 and their maximum values are $0.32$ for all the models, indicating that dust extinction 
does not work initially and the CO abundances are extremely low.
As the atomic gas accumulates into the post-shock layer, the mean value of $A_\mathrm{V,eff}$  
increases with time.
In addition, spatial variations of $A_\mathrm{eff}$ are enhanced by shock compression. 
Fig. \ref{fig:av} shows the PDFs of $A_\mathrm{eff}$ in the post-shock layers 
for models $\Theta3$, $\Theta11$, and $\Theta36$.
Around $t=1$~Myr, the high $A_\mathrm{V,eff}$ tails of the PDFs start to exceed $A_\mathrm{V,eff}=1$ 
for all the models (Fig. \ref{fig:av}). 
This indicates that the regions shielded by dust grains are formed at a epoch near $t\sim 1$~Myr.
At that epoch, $f_\mathrm{CO}$ begins to grow (the bottom-left panel of Fig. \ref{fig:H2}). 

Unlike the H$_2$ formation, the CO formation proceeds more rapidly for model $\Theta11$ 
than for models $\Theta3$ and $\Theta36$.
This is because the gas tends to have higher $A_\mathrm{V,eff}$ for model $\Theta11$ than 
for models $\Theta3$ and $\Theta36$ as shown in the PDFs at $t=5~$Myr in Fig. \ref{fig:av}.
In addition, model $\Theta11$ has a larger amount of the dense gas with $n>10^3~\pcc$ (Fig. \ref{fig:denhist}).  
An increase in gas density promotes the CO formation.

Note that the fraction of CO is still less than 20\% 
even for model $\Theta11$ because we terminate the simulations at the 
relatively early epoch of $t=5~$Myr to ignore self-gravity.
The CO formation is expected to proceed and a significant fraction of the gas 
will be fully molecular after $t=5~$Myr. 
The later evolution will be discussed in Section \ref{sec:discussion}.

\section{An Analytical Model Describing Time Evolution of the Post-shock Layers}\label{sec:ana}

Our results showed that there is a critical angle denoted by $\thetacr$ 
above which 
the shock-amplified magnetic field controls the post-shock layers. 
We define the critical angle $\thetacr$ as the angle above which 
the velocity dispersion parallel to the $x$-axis obeys $\delta v_{x0}(t/t_0)^{-1/2}$.
From Figs. \ref{fig:vdisp}a and \ref{fig:dvx_th}, the critical angle is set to $11^\circ$ 
in the fiducial parameter set.
In Sections \ref{sec:pre}-\ref{sec:meanden}, we found that 
the time evolution of the post-shock layer can be understood using the 
pressure balance between the pre- and post-shock gases.
In this section, we establish a simple analytic model that describes the global time evolution of 
the post-shock layers.

\subsection{Formulation}\label{sec:comp}
The pressure balance between the post- and pre-shock gases is given by 
\begin{equation}
     \Pturb_\mathrm{ana} + \Pmag_\mathrm{ana} \sim \rhoave V_0^2,
     \label{momx}
\end{equation}
where $\Pturb_\mathrm{ana} \equiv \rhosh \delta v_x^2$, 
$\Pmag_\mathrm{ana} = \Bshp^2/8\pi$, and 
$B_\mathrm{sh\perp}$ is the mean transverse field strength. 
The effect of the parallel field component is omitted in Equation (\ref{momx}) 
because $V_0$ is super-Alfv{\'e}nic.
The effect of the radiative cooling 
is implicitly considered by ignoring
the thermal pressure in Equation (\ref{momx}).
The contribution of the thermal pressure to the total momentum flux 
is negligible for all the models (Fig. \ref{fig:1dim_pre}).
The magnetic flux conservation across a shock front is given by 
\begin{equation}
     \frac{\Bshp}{\rhosh}
     = \frac{B_0\sin\theta}\rhoave.
     \label{flux}
\end{equation}
Combining Equation (\ref{momx}) with (\ref{flux}), one obtains 
\begin{equation}
     \frac{\rhosh}{\rhoave}
     = - \left(\frac{\delta v_x}{\CAp} \right)^2
     + \sqrt{ \left( \frac{\delta v_x}{\CAp} \right)^4 + 2\left(\frac{V_0}{\CAp}\right)^2}
     \label{rhosh}
\end{equation}
where $C_\mathrm{A0\perp} = B_0\sin\theta/\sqrt{4\pi \langle \rho_0\rangle}$
is the upstream Alfv{\'e}n speed with respect to the transverse component of the upstream magnetic field.

Equations (\ref{momx})-(\ref{rhosh}) imply that $\delta v_x$ is 
related to the post-shock structure.
The ram pressure becomes equal to the magnetic stress when 
the velocity dispersion satisfies $\delta v_x=\dvxeq$, where 
\begin{eqnarray}
     \dvxeq &=& \sqrt{\frac{V_0\CAp}{2}}\nonumber \\
     & = & 2.9~\mathrm{km~s^{-1}}~\left( \frac{V_0}{20~\mathrm{km~s^{-1}}} \right)^{1/2}
     \left( \frac{B_0\sin\theta}{5~\mu\mathrm{G}\times \sin 11^\circ} \right)^{1/2}\nonumber \\
     &&\hspace{3cm}
     \times \left(  \frac{\nave}{5~\pcc} \right)^{-1/4}
     \label{dvxeq}
\end{eqnarray}

If $\delta v_x \gg \dvxeq$, the ram pressure dominates 
over the magnetic stress.
In this case, Equation (\ref{rhosh}) is 
reduced to $\rhosh/\rhoave\sim \left(V_0/\delta v_x\right)^2$. 
If $\delta v_x \ll \dvxeq$, mainly the magnetic stress 
supports the post-shock layers. 
In this case, the mean post-shock density is given by 
\begin{equation}
     \frac{\rhosh}{\rhoave}\sim 
     \frac{\rhosh_\mathrm{m}}{\rhoave}\equiv 
     \frac{V_0 \sqrt{8\pi \rhoave}}{B_0\sin\theta}.
     \label{rhoshmag}
\end{equation}
Equation (\ref{rhoshmag}) was derived by \citet{McKee1980} and \citet{II2009}.

\begin{figure}[htpb]
\centering
\includegraphics[width=8cm]{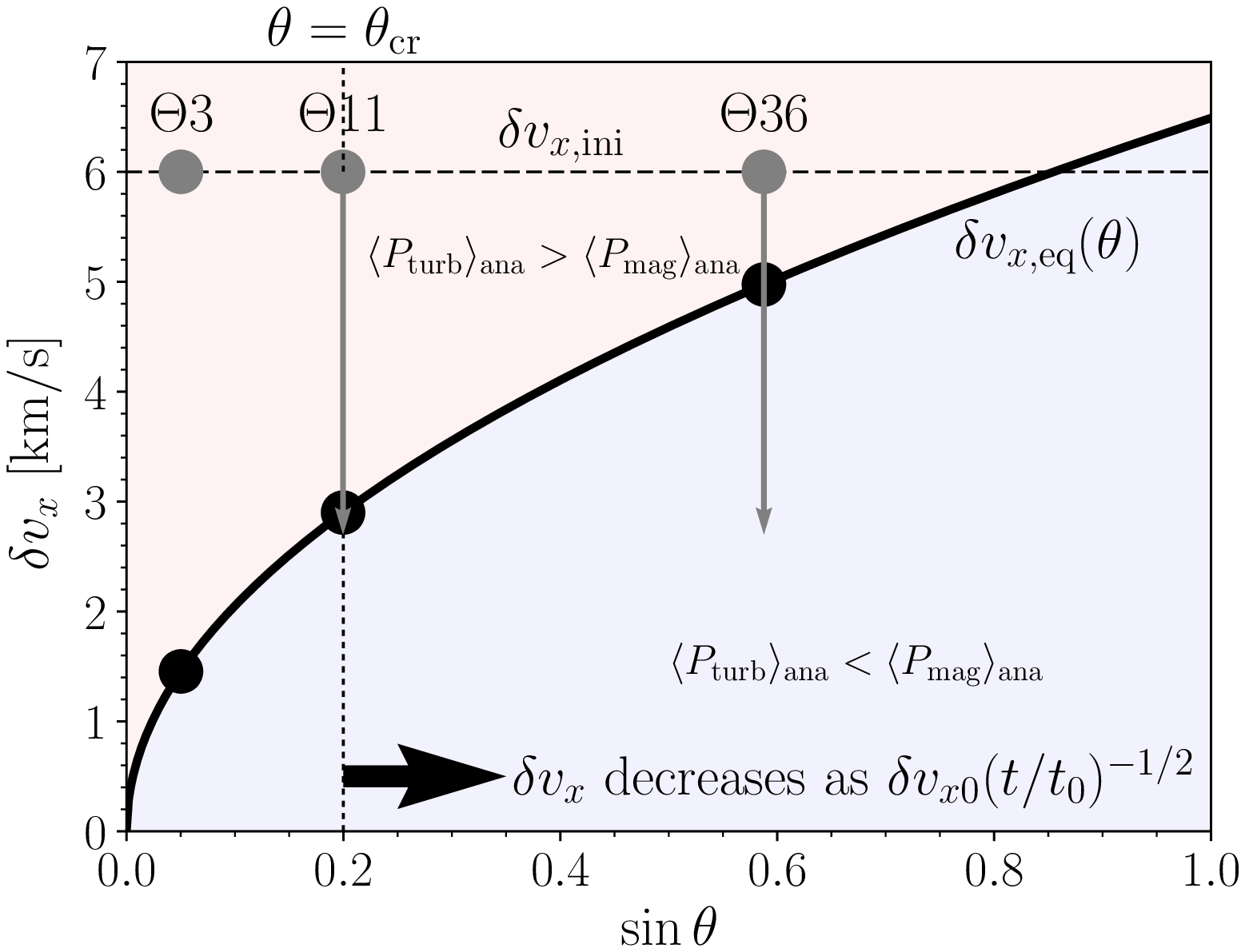}
\caption{
The critical velocity dispersion $\dvxeq$ for which 
the ram pressure is equal to the magnetic stress
as a function of $\sin\theta$ (black solid line).
The red and blue regions correspond to the $\Pturb_\mathrm{ana}$-dominated and $\Pmag_\mathrm{ana}$-dominated 
regions, respectively.
The velocity dispersion parallel to the $x$-axis in the very early phase is plotted as
the horizontal dashed line.
The dotted vertical line corresponds to the critical angle below which the turbulence is maintained.
The three gray circles indicate the velocity dispersions in the early 
phases for models $\Theta3$, $\Theta11$, and 
$\Theta36$.
The velocity dispersions giving equality for models $\Theta3$, $\Theta11$, and $\Theta36$ are 
shown by the three black circles.
The gray arrows represent the time evolutions of $\delta v_x$ for the models until $t=5~$Myr.
}
\label{fig:dvxeq}
\end{figure}

Fig. \ref{fig:dvxeq} shows $\dvxeq$ as a function of $\sin\theta$.
If $\delta v_x$ is larger (smaller) than $\dvxeq$, $\Pturb_\mathrm{ana}$ ($\Pmag_\mathrm{ana}$) 
dominates in the post-shock layers.
Let us consider the time evolution of the post-shock layers in this figure.
In the early phase, the super-Alfv{\'e}nic 
collision of the high density-inhomogeneous gas drives the 
longitudinal velocity dispersion as large as 
$\sim 6~$km~s$^{-1}$ (the horizontal dashed line), regardless 
of $\theta$ (Fig. \ref{fig:vdisp}).
The later evolution of $\delta v_x$ is determined by 
the critical angle of $\thetacr$ 
(the vertical dotted line).
If $\theta$ is less than $\thetacr$, $\delta v_x$ does not decrease with time 
and remains the initial value of $\sim 6$~km~s$^{-1}$.
As long as $\theta\ge \thetacr$, we found that $\delta v_x$ decreases as 
$v_{x0}(t/t_0)^{-1/2}$, regardless of $\theta$ 
as in Figs. \ref{fig:vdisp}a and \ref{fig:dvx_th}.  
The gray arrows indicate the time evolutions of $\delta v_x$ 
until $t=5$~Myr for models $\Theta11$ and $\Theta36$.
Fig. \ref{fig:dvxeq} clearly shows that the magnetic stress overtakes 
the ram pressure 
earlier for model $\Theta36$ than for model $\Theta11$ since 
the difference between $\delta v_x(t)$ and $\dvxeq$ indicates the significance of the ram pressure.

The analytic formula for the time evolution of $\delta v_x$ (Equation (\ref{vx}))
allows us to express Equations (\ref{momx})-(\ref{rhosh}) as a function of time.
At a given $\theta$, $\Pturb_\mathrm{ana}$ becomes equal to $\Pmag_\mathrm{ana}$,
and $\delta v_x$ reaches $\dvxeq$ when $t=t_\mathrm{eq}$, where 
\begin{eqnarray}
     t_\mathrm{eq} &=& 4.4~\mathrm{Myr} 
     \left( \frac{\delta v_{x0}/V_0}{0.3} \right)^2
     \left( \frac{V_0}{20~\mathrm{km~s^{-1}}} \right)\nonumber \\
     & & \hspace{1cm}
     \left( \frac{\nave}{5~\pcc} \right)^{1/2}
     \left( \frac{B_0\sin\theta}{5~\mu\mathrm{G}\times \sin11^\circ} \right)^{-1}.
     \label{teq}
\end{eqnarray}

\subsection{Comparison with the Simulation Results}\label{sec:comp}
\subsubsection{Mean Magnetic Stresses}\label{sec:premag}
\begin{figure}[htpb]
\centering
\includegraphics[width=8cm]{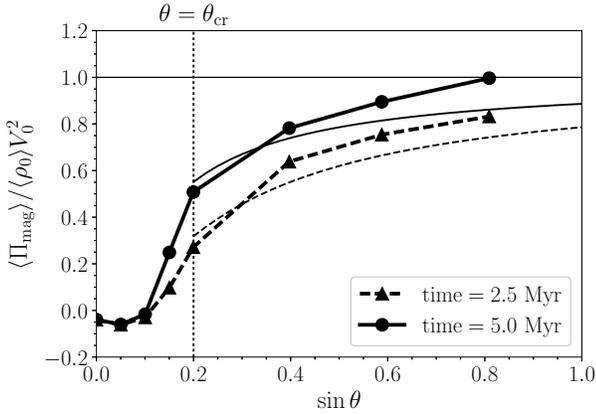}
\caption{
Dependence of $\Pmag$ on $\sin\theta$. 
The solid and dashed lines correspond to the magnetic stresses measured at $t=2.5$~Myr and $5.0$~Myr, 
respectively.
The vertical axis is normalized by the upstream ram pressure.
The dotted vertical line represents the critical angle.
}
     \label{fig:pre_th}
\end{figure}

Fig. \ref{fig:pre_th} shows $\Pmag$ 
measured at $t=2.5$~Myr and $t=5~$Myr as a function of $\sin\theta$.
The ram 
pressures are not plotted because $\Pturb+\Pmag \sim \rhoave V_0^2$ is satisfied.

The behavior of $\Pmag$ is determined by whether $\theta$ is larger than $\thetacr$ or not.
If $\theta<\thetacr$, $\Pmag$ is almost independent of time and remains almost zero.
For $\theta>\thetacr$, $\Pmag$ approaches $\langle \rho_0 \rangle V_0^2$ as turbulence decays.
The magnetic stresses 
for larger $\theta$ models reach $\langle \rho_0 \rangle V_0^2$ 
earlier as shown in Fig. \ref{fig:dvxeq}.
At each epoch, 
the analytic estimate of the magnetic stress is plotted as the thin line.
It is confirmed that 
the predictions from the analytic model are consistent with the simulation results, although there are 
some discrepancies.
The analytic model slightly underestimates the magnetic stress 
for larger $\theta$ because 
of the weak negative dependence of $\delta v_x$ on $\theta$ found 
in Figs. \ref{fig:vdisp}a and \ref{fig:dvx_th}.

We should note that, strictly speaking, $\Pmag$ remains almost zero
not for $\theta<\thetacr$ but for $\theta<0.6\thetacr~(\sin \theta<0.12)$ in 
Fig. \ref{fig:pre_th}.
A model with $\theta\le 0.6\thetacr $ does not show a decrease in $\Pturb$.
For $\theta\ge \thetacr$, the shock-amplified magnetic field controls the post-shock dynamics and 
$\delta v_x$ decreases, obeying $\delta v_x \sim \delta v_{x0} (t/t_0)^{-1/2}$.
The angle range of $0.6\thetacr<\theta<\thetacr$ exhibits 
the transition between the layers regulated by ram pressure 
($\Pmag\sim 0,~\Pturb \sim \rhoave V_0^2$) and 
those regulated by magnetic stress ($\delta v_x \sim \delta v_{x0}(t/t_0)^{-1/2}$).
Although the importance of $\Pmag$ increases with time 
in the total momentum flux, the ram pressure disturbs the post-shock layer significantly, 
leading to a slower decrease in $\delta v_x$ ($\delta v_x > \delta v_{x0}(t/t_0)^{-1/2}$).


\subsubsection{Mean Post-shock Densities}
\begin{figure}[htpb]
     \centering
     \includegraphics[width=8cm]{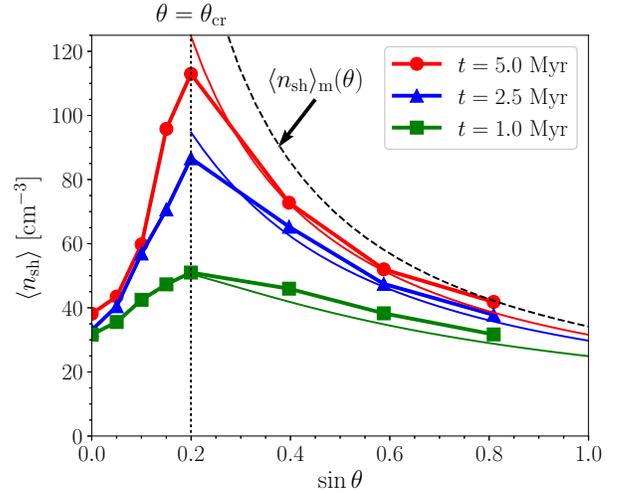}
     \caption{
     The mean post-shock densities at the three different epochs 
     $t=5~$Myr (red), $t=2.5$~Myr (blue), and $t=1$~Myr (green) are plotted 
     as a function of $\sin \theta$. 
     At each epoch, the analytic estimation of $\nsh$ (Equation (\ref{rhosh})) is 
     plotted as the thin line with the same color.
     The dashed line indicates $\nsh_\mathrm{sh}$ given by Equation (\ref{rhoshmag}).
     The dotted vertical line represents the critical angle.
     }
     \label{fig:denave}
\end{figure}

Fig. \ref{fig:denave} shows $\langle n_\mathrm{sh}\rangle$ measured at three different epochs 
as a function of $\sin\theta$.
The dashed line corresponds to $\nsh_\mathrm{m}$ as a function of $\sin \theta$ (Equation (\ref{rhosh})).
At a fixed $\theta$ 
the mean post-shock density approaches $\rhosh_\mathrm{m}$ as $\delta v_x$ decreases,
because the difference between $\nsh$ and $\nsh_\mathrm{m}$ 
indicates the significance of the ram pressure. 
The analytic estimates shown in Equation (\ref{rhosh}) at the three different epochs are plotted 
by the three thin lines.
The mean post-shock densities predicted from Equation (\ref{rhosh}) 
are consistent with those derived from the simulation results at each epoch for $\theta\ge \thetacr$.

Note that the angle ($\theta\sim \thetacr$)
that gives the maximum $\nsh$ does not depend on time, although $\nsh$ for larger $\theta$ 
reaches $\nsh_\mathrm{m}$ earlier.
This is because $\nsh_\mathrm{m}\propto (\sin\theta)^{-1}$ is a decreasing function of $\theta$.

Although the magnetic stress starts to show a considerable growth at $ \theta = 0.6 \thetacr$ 
in Fig. \ref{fig:pre_th}, $\nsh$ appears to increase smoothly for $\theta < \thetacr$ 
in Fig. \ref{fig:denave}.
The smooth increase of $\nsh$ for $\theta\le 0.6\thetacr$ is explained as follows.
For $\theta\le 0.6\thetacr$, $\delta v_x$ gradually decreases with $\theta$ (Fig. \ref{fig:dvx_th})
although $\Pturb \sim \rhoave V_0^2$ is constant.
From the relation $\Pturb \propto \nsh \delta v_x^2$, $\nsh$ should increase with $\theta$ 
as shown in Fig. \ref{fig:denave}.
Although $\nsh$ increases smoothly with $\theta$ for $\theta< \thetacr$, 
its manner of the time evolution of $\nsh$ suddenly changes at an angle near $\theta = 0.6 \thetacr$. 
Fig. \ref{fig:denave} clearly shows that $\nsh$ increases with time for 
$\sin \theta = 0.15~(\theta = 0.75\thetacr)$ 
while $\nsh$ does not increase significantly and is saturated around $60~\pcc$ for 
$\sin \theta = 0.1~(\theta=0.5\thetacr)$.

\begin{figure}[htpb]
     \centering
     \includegraphics[width=8cm]{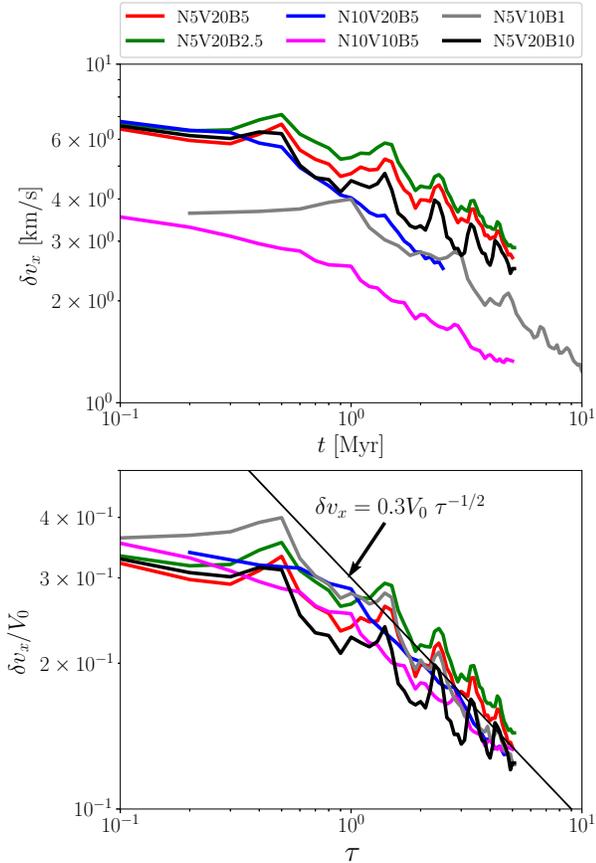}
\caption{
Top panel: time evolution of the longitudinal velocity dispersions for various models.
The velocity dispersions are measured for 
N5V20B5 ($\theta=11^\circ$), N10V20B5 ($\theta=23^\circ$),  
N5V10B1 ($\theta=36^\circ$), N5V20B2.5 ($\theta=23^\circ$), 
N10V10B5 ($\theta=11^\circ$), and N5V20B10 ($\theta=11^\circ$). 
The bottom panel is the same as the top panel but the horizontal axis is changed to 
$\tau=t~(\nave/5~\pcc)(V_0/20~\mathrm{km~s^{-1}})/(1~\mathrm{Myr})$ 
and the vertical axis is normalized by $V_0$.
}
     \label{fig:veldisp_para}
\end{figure}

\begin{figure}[htpb]
     \centering
     \includegraphics[width=8cm]{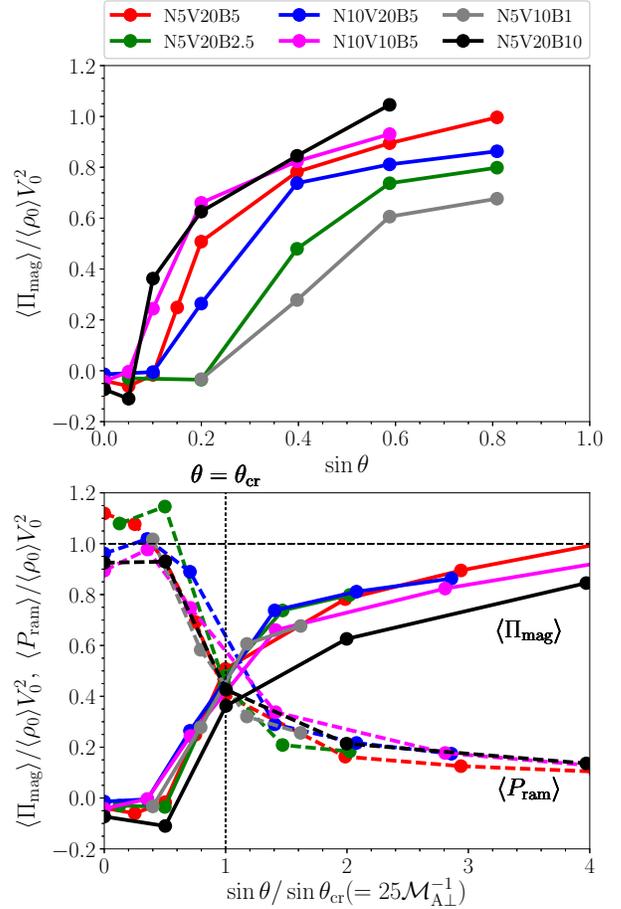}
\caption{
Top panel: mean magnetic stresses in the post-shock layers for various parameter sets 
of $(\langle n_0\rangle, V_0,  B_0)$, which are tabulated in Table \ref{tab},
as a function of $\sin \theta$.
The magnetic stresses are measured at $\tau=5$.
The bottom panel is the same as the top panel but 
the horizontal axis 
is normalized by the critical value ($\sin \theta_\mathrm{crit}$) shown in Equation (\ref{th_crit}).
For each model, 
the mean ram pressures in the post-shock layer are also plotted as the dashed line with the same color.
The vertical dotted line represents the $\theta=\thetacr$ line.
}
     \label{fig:force}
\end{figure}
\section{Parameter Survey}\label{sec:parameter}
In Section \ref{sec:results}, we presented the results for
the fiducial parameter set ($\nave = 5~\mathrm{cm}^{-3}$, $B_0=5~\mu\mathrm{G}$,
$V_0=20~\mathrm{km~s^{-1}}$).
In this section, a parameter survey is performed by changing 
$(\nave,V_0,B_0)$. The adopted parameters are summarized in Table \ref{tab}.
To save computational costs, we 
conducted the parameter survey with 
half the resolution, $512\times 256\times 256$, compared to the fiducial parameter set shown
in Section \ref{sec:results}.
We have checked that at least the global quantities, $\Pturb$ and $\Pmag$, 
are consistent with those with twice the resolution (the fractional differences are as small as $10\%$ in 
the fiducial parameter set).
In each of the models, the simulations are performed by changing $\theta$, and  
$\Pturb$ and $\Pmag$ are calculated as functions of $\theta$.

The top panel of Fig. \ref{fig:veldisp_para} shows the time evolution of $\delta v_x$
for various models.
For each model, we measure $\delta v_x$ at an angle $\theta$ where $\Pmag$ increases with time.
The longitudinal velocity dispersions for all the models decrease with time 
in a similar manner
(Fig. \ref{fig:vdisp}a).
The bottom panel of Fig. \ref{fig:veldisp_para} shows that 
$\delta v_x$ roughly follows a universal law, 
\begin{equation}
        \delta v_x = 0.3 V_0 \tau^{-1/2}.
        \label{dvx_para}
\end{equation}
where $\tau=t\times (\nave/5~\pcc)\left( V_0/20~\mathrm{km~s^{-1}}\right)/(1~\mathrm{Myr})$.
Equation (\ref{dvx_para}) is reduced to Equation (\ref{vx}) in the fiducial model.
Interestingly, the time evolution of $\delta v_x/V_0$ does not depend on the 
field strength sensitively, 
and it is characterized only by the accumulated mean column density 
$\sigma = 2\rhoave V_0 t$, which is proportional to $\tau$.
We compare these results for the models at $\tau=5$ 
when the mean visual extinction reaches 1.6.
\begin{table}
     \begin{tabular}{|c||c|c|c|c|}
             \hline
             Model name & 
             \begin{minipage}{1cm}
             \vspace{1mm}
                 \centering
             $\nave$ 

             \vspace{1.0mm}

             [cm$^{-3}$]

             \vspace{1.0mm}
             \end{minipage}
             & 
             \begin{minipage}{1cm}
             \vspace{1.0mm}
                 \centering
             $V_0$

             \vspace{1.0mm}

             [km~s$^{-1}$] 

             \vspace{1.0mm}
             \end{minipage}
             & 
             \begin{minipage}{1cm}
             \vspace{1.0mm}
                 \centering
             $B_0$ 

             \vspace{1.0mm}

             [$\mu$G] 

             \vspace{1.0mm}
             \end{minipage}
             & 
             \begin{minipage}{1cm}
                     \centering
             $\displaystyle 
             {\cal M}_\mathrm{A}$
             \end{minipage}
             \\
             \hline
             N5V20B5   & 5 & 20 & 5 & 4.9\\
             N5V20B2.5 & 5 & 20 & 2.5 & 9.7\\
             N5V10B1   & 5 & 10 & 1 & 12\\
             N10V20B5  & 10 & 20 & 5 & 6.9\\
             N10V10B5  & 10 & 10 & 5 & 3.4\\
             N5V20B10  & 5 & 20 & 10 & 2.4\\
             \hline
     \end{tabular}
     \caption{
     Model parameters. The last column indicates the Alfv{\'e}n Mach number 
     ${\cal M}_\mathrm{A}=V_0 \sqrt{4\pi \rhoave}/B_0$}
     \label{tab}
\end{table}

The top panel of Fig. \ref{fig:force} 
shows $\Pmag$ as a function of $\sin\theta$ for various parameter sets.
For a given parameter set $(\langle n_0\rangle, V_0, B_0)$, the $\theta$-dependence of $\Pmag$ 
is similar to that for the fiducial parameter set as shown in Fig. \ref{fig:pre_th}.
The averaged magnetic stress $\Pmag$ increases monotonically with $\theta$.
The only difference is the values of the critical angles below which $\Pmag$ is almost zero.

Here, we present an analytic estimate of the critical angle that
can explain the results in different parameter sets ($\nave$, $V_0$, $B_0$)
using the analytic model developed in Section \ref{sec:ana}.
We found that $\delta v_x$ is roughly proportional to $V_0$ at least in the 
$V_0$ range $10~\mathrm{km~s^{-1}}\le V_0 \le 20~\mathrm{km~s^{-1}}$ 
for large angles from Fig. \ref{fig:veldisp_para}. At the critical angle, 
$\dvxeq(\thetacr)$ is equal to $0.15V_0$
in the fiducial parameter set (Fig. \ref{fig:dvxeq}).
If this is the case also for other parameter sets, one obtains 
\begin{equation}
        \sin \thetacr = 0.2
     \left( \frac{\nave}{5~\mathrm{cm}^{-3}} \right)^{1/2}
     \left(  \frac{V_0}{20~\mathrm{km~s^{-1}}} \right)
     \left(  \frac{B_0}{5~\mu\mathrm{G}} \right)^{-1}.
     \label{th_crit}
\end{equation}
Equation (\ref{th_crit}) is rewritten as 
\begin{equation}
     {\cal M}_\mathrm{A\perp,cr}= \frac{V_0}{\CAp(\theta_\mathrm{cr})} = 25,
     \label{macha}
\end{equation}
meaning that if the Alfv{\'e}n Mach number with respect to the perpendicular field component,
${\cal M}_\mathrm{A\perp}=V_0/\CAp$, is larger than 25, 
the super-Alfv{\'e}nic velocity dispersion is maintained.

Let us derive the critical angle (Equation (\ref{th_crit})) from the following simple argument.
In the early phase, the velocity dispersion parallel to the $x$-axis 
takes a roughly constant value of
\begin{equation}
\delta v_x \sim \sqrt{\epsilon} V_0,
\label{dvx}
\end{equation}
where Equation (\ref{eps}) is used.
If the magnetic stress mainly supports the post-shock layer against the upstream ram pressure, 
the post-shock Alfv\'en speed $C_\mathrm{A,sh}$ becomes
\begin{equation}
	C_\mathrm{A,sh} = \left( \frac{ V_0 B \sin \theta }{\sqrt{2\pi \rhoave}}\right)^{1/2},
\label{CAsh}
\end{equation}
where we use Equation (\ref{rhoshmag}) and $B_\mathrm{sh,\perp}^2/8\pi = \rhoave V_0^2$.
If $\delta v_x < C_\mathrm{A,sh}$, the magnetic field cannot be bent by the 
velocity dispersion parallel to the $x$-axis.
From Equations (\ref{dvx}) and (\ref{CAsh}), the critical angle $\thetacr'$ satisfying 
$\delta v_x \sim C_\mathrm{A,sh}$ is given by 
\begin{equation}
\sin \thetacr' \sim 0.3 \left( \frac{\epsilon}{0.1} \right)
     \left( \frac{\nave}{5~\mathrm{cm}^{-3}} \right)^{1/2}
     \left(  \frac{V_0}{20~\mathrm{km~s^{-1}}} \right)
     \left(  \frac{B_0}{5~\mu\mathrm{G}} \right)^{-1}.
     \label{th_critd}
\end{equation}
The parameter dependence of $\sin\thetacr'$ is consistent with that of $\sin\thetacr$
although $\sin \thetacr'$ is slightly larger than $\sin \thetacr$.

The bottom panel of Fig. \ref{fig:force} clearly shows that 
each of $\Pmag/\rhoave V_0^2$ and $\Pturb/\rhoave V_0^2$ 
for all the models follows a universal line if
$\sin\theta/\sin\thetacr$ is used as the horizontal axis.
The ratio $\sin \theta/\sin \thetacr$ can be expressed as 
$25{\cal M}_\mathrm{A\perp}^{-1}$.
The reason why both $\Pmag/\rhoave V_0^2$ and $\Pturb/\rhoave V_0^2$ 
depend only on ${\cal M}_\mathrm{A\perp}$ 
is that all the models have almost the same $\delta v_x/V_0\sim 0.13$ at $\tau=5$ 
(Fig. \ref{fig:veldisp_para}).
From Equation (\ref{rhosh}), one finds that $\rhosh/\rhoave$ depends only 
on ${\cal M}_\mathrm{A\perp}$ 
if $\delta v_x/V_0$ is fixed at 0.13.
Substituting Equation (\ref{rhosh}) with $\delta v_x/V_0=0.13$ into 
Equations (\ref{momx}) and (\ref{flux}), 
it is found that $\Pmag/\rhoave V_0^2$ and $\Pturb/\rhoave V_0^2$ are
determined only by ${\cal M}_\mathrm{A\perp}$ as in  $\rhosh/\rhoave$.

\begin{figure}[htpb]
     \centering
     \includegraphics[width=8.5cm]{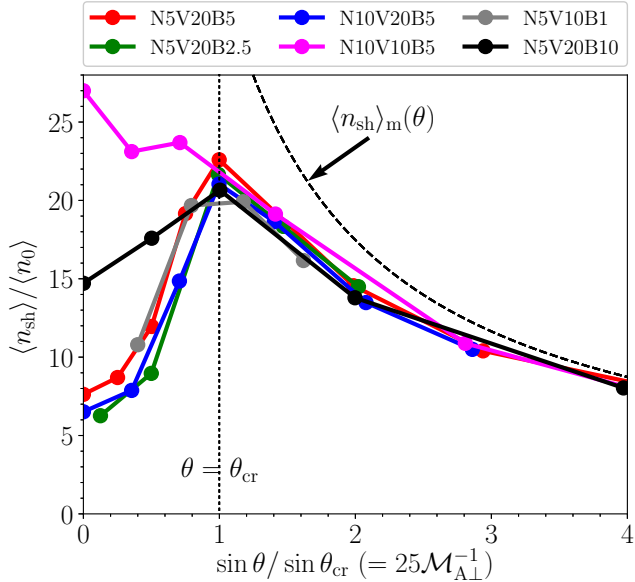}
\caption{
Mean post-shock densities for various parameter sets 
of $(\langle n_0\rangle, V_0,  B_0)$, which are summarized in Table \ref{tab},
as a function of $\sin \theta$.
The post-shock densities are measured at 
$t=5~\mathrm{Myr}~(\nave/5~\pcc)^{-1}\left( V_0/20~\mathrm{km~s^{-1}} \right)^{-1}$.
The vertical axis is normalized by the upstream mean density and 
the horizontal axis 
is normalized by the critical value ($\sin \theta_\mathrm{crit}$) shown in Equation (\ref{th_crit}).
The dashed line corresponds to $\nsh_\mathrm{m}$ shown in Equation (\ref{rhoshmag}).
The vertical dotted line shows $\theta=\thetacr$.
}
     \label{fig:force_denave}
\end{figure}

The mean post-shock densities normalized by $\nave$
for various models are plotted 
as a function of $\sin \theta/\sin\thetacr$ in Fig. \ref{fig:force_denave}. 
The dashed line corresponds to $\nsh_\mathrm{m}$, which is rewritten 
as $\nsh_\mathrm{m}/\nave = 35~(\sin \theta/\sin\thetacr)^{-1}$.
For angles larger than $\sim \thetacr$,
$\nsh/\nave$ in all the models is approximated by an universal line as a function of $\sin\theta/\sin\thetacr$ 
(Fig. \ref{fig:force_denave}).
By contrast,
the behavior of $\nsh/\nave$ for angles smaller than $\sim \thetacr$
can be divided into two groups.
One group contains models N10V20B5, N5V10B1, and N5V20B2.5, where 
$\nsh$ rapidly decreases with decreasing $\theta$ 
from $\theta\sim \thetacr$ in a manner similar to that for the fiducial parameter set (N5V20B5).
This group is referred to as the low-$\nsh$ group.
All the lines of this group converge to a universal line in the 
($\sin\theta/\sin\thetacr$, $\nsh/\nave$) plane.
However, models N10V10B5 and N5V20B10 do not show a rapid decrease in $\nsh$ 
as $\theta$ decreases.
Especially for model N10V10B5, the post-shock layers are dense even for $\theta<\thetacr$,
and a sharp $\nsh$ peak disappears.

The density enhancement for $\theta<\thetacr$ in model N10V10B5 arises from 
the fact that $\delta v_x$ gradually decreases with time. 
In the low-$\nsh$ group including the fiducial model, 
$\delta v_x$ is found not to decrease for $\theta<\thetacr$ because 
the CNM clumps are not decelerated significantly 
after passing through the shock fronts. 
We speculate that 
the difference between the $\delta v_x$ evolutions 
for model N10V10B5 and the low-$\nsh$ group comes from the following three points.
First, because 
${\cal M}_\mathrm{A}$ is lower for model N10V10B5 than for the low-$\nsh$ group 
where ${\cal M}_\mathrm{A}\gtrsim 5$ (Table \ref{tab}),
the magnetic field is expected to work more effectively 
than in the low-$\nsh$ group.
Indeed, at $\tau = 5$, $\Pmag/\rhoave V_0^2$ for model N10V10B5 is 2.5 times larger than 
that for the fiducial model.
However, lower Alfv{\'e}n Mach numbers are not a sufficient condition to get larger $\nsh/\nave$
because $\nsh/\nave$ is larger for model N10V10B5 than for model N5V20B5 although 
${\cal M}_\mathrm{A}$ is lower for model N5V20B5.
Thus, the other points appear to be required.
The second point is that the collision speed for model N10V10B5 is lower than in the fiducial model,
indicating that the upstream CNM clumps have lower momenta, which 
allow them to be decelerated more easily. 
The third point is 
the larger upstream mean density for model N10V10B5, which enhances the volume filling 
factor of the CNM phase. A collision between CNM clumps is more probable than 
one between a CNM clump and WNM.

We should note that even for models N10V10B5 and N5V20B10 
the super-Alfv{\'e}nic turbulence 
is maintained without decay because $\Pturb$ is much larger than $\Pmag$ as shown 
in Fig. \ref{fig:force}.
The decrease in $\delta v_x$ does not lead to an increase in $\Pmag$.
This contrasts 
with the case with $\theta>\thetacr$, in which $\Pmag$ increases with decreasing $\delta v_x$.
For $\theta<\thetacr$, the decrease in $\delta v_x$
is compensated by the increase in the mean post-shock density 
to maintain pressure balance between $\Pturb$ and the upstream ram pressure.

\section{Discussion}\label{sec:discussion}
\subsection{Sub-Alfv{\'e}nic Colliding Flows}\label{sec:sub}
The parameter survey in Section \ref{sec:parameter} 
shows that the global time evolution of the post-shock layers 
is described by the analytic model.
It should be noted that the simulation results 
depend not on the parallel field component ($B_0\cos\theta$) but only on the perpendicular field 
component ($B_0\sin\theta$).
This is simply because super-Alfv{\'e}nic colliding flows are considered in this paper.
The parallel field component does not play an important role.

The analytic model thus cannot be applied for all parameter sets of ($\nave,V_0,B_0$).
If a colliding flow is sub-Alfv{\'e}nic,
the magnetic field is too strong to be bent by shock compression.
Instead, the gas is allowed to accumulate along the magnetic field through slow shocks. 
From the MHD Rankine-Hugoniot relations, 
\citet{II2009} derived a criterion required to form slow shocks as follows:
\begin{equation}
     B_0 > \Bcr \equiv \frac{2}{3}V_0 \sqrt{4\pi \rho_0},
     \label{Bcr}
\end{equation}
where $\rho_0$ is the uniform upstream density and the ratio of specific heats is set to $5/3$.
We should note that equation (\ref{Bcr}) is different from equation (14) in \citet{II2009}, where
the left-hand side of equation (\ref{Bcr}), $B_0$, is replaced by $B_0\cos\theta$.
Equation (\ref{Bcr}) corresponds to a necessary condition to form slow shocks.
If equation (\ref{Bcr}) is not satisfied, slow shocks are not generated for any angle. 
Equation (\ref{Bcr}) is roughly the same as the condition 
of sub-Alfv\'enic colliding flows, which is given by 
$B_0> V_0\sqrt{4\pi\rho_0}$.

\begin{figure}[htpb]
     \centering
     \includegraphics[width=8cm]{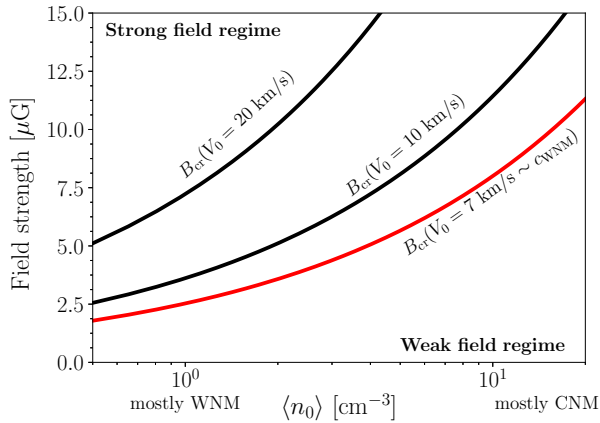}
     \caption{
        Density dependence of $B_\mathrm{cr}$ for three different collision speeds.
        At each collision speed, the region above (below) the line $B=\Bcr$ 
        belongs to the strong (weak) field 
       regime.
     }
     \label{fig:weakstrongB}
\end{figure}

Fig. \ref{fig:weakstrongB} shows $\Bcr$ as a function of $\nave$ for three different 
collision speeds. In the figure, for a given $V_0$, the region above (below) the line $B=\Bcr$ is 
referred to as the strong (weak) field regime.
The range of $B_0$ that belongs to the weak field regime is wider for higher mean densities because 
Alfv{\'e}n speed is a decreasing function of gas density.
Shock compression of denser atomic gases is likely to be in the weak field regime,
where the analytic model can be applied.
Since this paper focuses on the MC formation from dense atomic gases that have already been 
piled up by previous episodes of compression, all our simulations are in the weak field regime.

\subsection{Comparison with Previous Studies}\label{sec:comparison}

Most simulations of WNM colliding flow have been done in the cases where 
the collision direction is parallel to the mean magnetic field 
\citep{Henn2008,Banerjee2009,Vaz2011,Kortgen2015,Zam2018}.
One of the differences between two-phase and WNM colliding flows is that
the two-phase colliding flows are highly inhomogeneous.
Our results showed that the inhomogeneity of colliding flows
enhances longitudinal velocity dispersion for $\theta\sim 0^\circ$ 
\citep[also see][]{Inoue2012,Carroll2014,Forgan2018}.
We also found density enhancement for $\theta\sim 0^\circ$
in the models where the field strength is relatively large and 
the Alfv\'en Mach number of the colliding flow is close to unity in Fig. \ref{fig:denave}.
This is consistent with the results of 
\citet{Heitsch2009} and \citet{Zam2018} 
who found that the post-shock layers become denser for stronger magnetic fields 
with a fixed collision speed
\citep[also see][for isothermal colliding flows]{Heitsch2007}.

In the two-phase colliding flows, 
the pre-existing upstream CNM clumps become an ingredient of the post-shock CNM clumps.
For WNM colliding flows with oblique fields, 
\citet{II2009} showed that only HI clouds with a density of $\sim 10$~cm$^{-3}$ form.
\citet{Heitsch2009} also reported that the formation of dense clouds is prohibited 
in a WNM colliding flow with a perpendicular magnetic field.
Our results, however, show that a
large amount of the dense gas with $>10^2~\pcc$ exists
in the post-shock layers even for model $\Theta36$ (Fig. \ref{fig:mass}). 
This is simply because the dense gases with $>10^2~\pcc$ can be directly supplied by 
the accretion of the pre-existing CNM clumps, whose densities are enhanced over $10^2~\pcc$ by 
the super-Alfv{\'e}nic shock compression.
Note that the CNM mass fraction in the post-shock layer is 
larger than the upstream CNM mass fraction of $\sim 0.5$. 
This indicates that the dense gases with $>10^2~\pcc$ are also provided from the surrounding diffuse gas 
through the thermal instability.



\subsection{Implications for the Formation of MCs}\label{sec:mol}
We found that 
the CO molecules form efficiently around $\theta\sim \thetacr$ where 
the dense gas with $>10^3~\pcc$ is efficiently generated (Fig. \ref{fig:mass}).
For $\theta<\thetacr$ ($\theta>\thetacr$), 
the large anisotropic velocity dispersion (the magnetic stress) prevents
the gas from being dense enough to form CO molecules (Figs. \ref{fig:mass} and \ref{fig:H2}).
Although the simulations are terminated in the early phase, 
the mass fraction of CO-forming gases exceeds 
17\% for $\theta\sim \thetacr$.
The CO fraction will continue to increase if we follow further evolution of the post-shock layers.

The setup of head-on colliding flows of the atomic gas leads to
somewhat artificial results, especially for $\theta<\thetacr$ 
in the models belonging to the low-$\nsh$ group, or model $\Theta3$ for the fiducial model.
The lower post-shock mean density and faster shock propagation velocity for model $\Theta3$ 
can be explained as follows: 
the upstream CNM clumps that are not decelerated at the shock due to high density continue to stream 
roughly along the $x$-axis and finally hit/push the shock wave on the opposite side.  
We think that this effect can be expected only for a very limited astronomical situation where 
two identical flows collide as in the present simulations.  
If we consider, for instance, the growth of an MC through an interaction with 
a shocked HI shell created by a supernova shock or an expanding HII region \citep[e.g.][]{Inutsuka2015}, 
the interaction would not 
destroy the MC even if the HI shell accretes to the MC
along the magnetic field. This is because the MC would have a mass (or column density) 
enough to decelerate the accreting HI gas.  In addition to this, when we consider the effect of gravity, 
we can expect that the expansion of the shocked region for model $\Theta3$ is stopped around $t\gtrsim5$ Myr 
(see equation \ref{eq:tg}), 
because the freely flying CNM clumps leaving the shocked region are decelerated 
by the gravity (Appendix \ref{app:grav}).  In the previous simulation done by \citet{Inoue2012}, 
they prevented the free propagation of CNM that crosses the $x$-boundary planes by setting a viscosity at 
the boundaries to mimic the effect of gravity. They found that even for the condition $\theta=0^{\circ}$, 
a realistic MC can be formed at $t\sim7$ Myr.
Self-gravity also contributes to local gas compression.
This will further promote the formation of MCs.
By contrast, 
if $\theta$ is sufficiently large, the formation of MCs is expected 
to remain inefficient even 
if the self-gravity becomes important because density enhancement is 
suppressed by the magnetic pressure.

\begin{figure}[htpb]
     \centering
     \includegraphics[width=8cm]{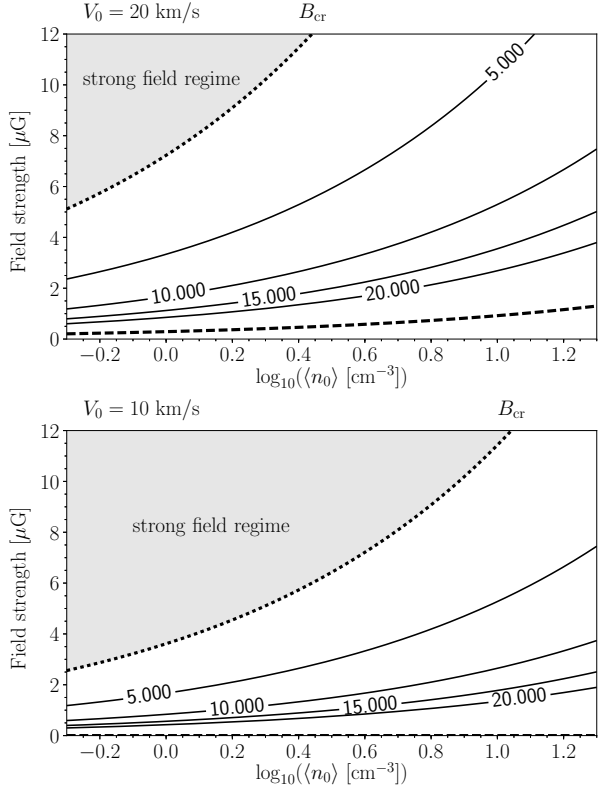}
     \caption{
             Contours of the critical angles in the ($\nave$, $B_0$) plane 
             for $V_0=20~$km~s$^{-1}$ (top) and $V_0=10~$km~s$^{-1}$ (bottom).
             The numbers on the contours indicate the corresponding angles in degrees.
             In each panel, the dotted line corresponds to the critical field strength 
             shown in equation (\ref{Bcr}). 
             In the regions below the dashed lines, the critical angles become $90^\circ$, indicating that 
             turbulence does not decay for any angle because of the low field strength.
     }
     \label{fig:thcrit}
\end{figure}

Our results show that the physical properties of cold clouds depend strongly on 
the field orientation especially around $\theta\sim \thetacr$.
To estimate how rare a shock compression with $\theta<\thetacr$ is,
the contours of the critical angles in the ($\nave,B_0$) plane for two different collision speeds are 
plotted in Fig. \ref{fig:thcrit}.
The gray regions in Fig. \ref{fig:thcrit} are not focused on in this paper 
(Section \ref{sec:sub}).

Fig. \ref{fig:thcrit} shows that the critical angles are very small unless magnetic fields 
are weak ($B_0<1~\mu$G).
To estimate the probability of realizing a shock compression with $\theta \le \thetacr$ ($P(\theta\le \thetacr)$),
it is assumed that a shock compression occurs isotropically in a uniform magnetic field.
The probability is proportional 
to the solid angle $\Omega$ of a cone with apex angle $2\thetacr$ ($\Omega = 4\pi \sin^2(\thetacr/2)$).
Using $\Omega$, the probability is given by $P(\theta\le\thetacr) = 2\times \Omega/4\pi = 2\sin^2 (\thetacr/2)$,
where the factor of two is to account for the cases of $\theta\le \thetacr$ and $\theta\ge \pi - \thetacr$.
Finally, we obtain $P(\theta\le \thetacr) \simeq \thetacr^2/2$ because the critical angle is small 
for $B_0>1~\mu$G (Fig. \ref{fig:thcrit})
\citep[also see][]{Inutsuka2015}.
The probability $P(\theta\le \thetacr) = 0.017~(\thetacr/11^\circ)^2$ is extremely low in the fiducial parameter set.
From Fig. \ref{fig:H2}, shock compression with larger $\theta$ significantly suppresses
the formation of dense gases and CO.
At least in the early evolution of MCs, the 
shock compression with angles less than $\xi\thetacr$ ($\xi=1-2$) is expected to 
contribute to an efficient MC formation because $f_{>10^3}$ and $f_\mathrm{CO}$ decreases rapidly with $\theta$ 
for $\theta>\thetacr$ (Figs. \ref{fig:mass} and \ref{fig:H2}). 
The probability of realizing the compression with $\theta<\xi\thetacr$ is 
$P(\theta<\xi\thetacr) = 2\pi \xi^2\thetacr^2/4\pi \sim 
0.04~(\xi/1.5)^2(\thetacr/11^\circ)^2$.

The free parameter $\xi$ cannot be determined from our work because 
we focused on the early stage of the MC formation.
We expect that $\xi$ is determined as a function of ($\nave$, $V_0$, $B_0$) 
by whether cold clouds become massive enough 
for star formation to occur against the magnetic pressure.
We will present the results of long-term simulations including gravity in a future paper.

\subsection{Caveats}\label{sec:caveats}
Although the simple setup of steady head-on colliding flows enables us to 
investigate the detailed post-shock dynamics,
we ignore several aspects of physics that are important in the galactic environment.

\subsubsection{Head-on Collision}

We consider that colliding flows are generated 
by adjacent expanding super-bubbles and galactic spiral waves.
However, head-on collisions considered in this paper are an extreme case.
In realistic situations, we need to take into account an offset colliding flow,
which would induce a global shear motion in the post-shock layer.
The shear motion can affect
the post-shock dynamics significantly \citep{Fogerty2016}.

Colliding flows are not an optimal approximation in the cases of 
a single SN or multiple SNe, which are expected to contribute to the MC formation.
They generate a hot bubble that drives an expanding shell confined by 
the shock front on the leading side and the contact discontinuity, which separates the shell from
the hot bubble, on the trailing side. 
Especially for the almost parallel field models (for instance model $\Theta3$), 
the existence of two shock fronts strongly affects the post-shock dynamics 
as discussed in Section \ref{sec:mol}.
\citet{Kim2015} investigated 
the evolution of shells expanding into realistic two-phase atomic gases. 
Because they ignore magnetic fields, 
their setups relate to the almost parallel field models in this work. 
In their simulations, 
pre-existing HI clouds strongly deform both the shock front and contact discontinuity, 
and the shells are widened.
These behaviors are qualitatively consistent with our results. 
However, the effect of the contact discontinuity 
should affect the post-shock dynamics.

\subsubsection{Local Simulations}
\label{sec:comp}

Although the well-controlled setups of our local simulations allow us to 
perform a detailed analysis, 
the locality of our simulations is one of the caveats. 
In this paper, we consider a continuous supply of the atomic gas assuming constant 
values of $V_0$, $B_0$, and $\theta$. 

First, we should note that shock compression has a finite duration 
in the galactic environment.
We consider that colliding flows are generated 
by adjacent expanding super-bubbles and galactic spiral waves.
Although our simulations are terminated earlier than the typical duration of super-bubbles and
spiral waves (a few tens of megayears), it is worth discussing the time evolution of a post-shock layer 
after shock compression ceases.
In local simulations,
many studies took into account the finite duration of colliding flows
\citep{Vaz2007,Banerjee2009,Vaz2010,Vaz2011,Kortgen2015,Zam2018}.
As the shock compression weakens, $B_\mathrm{cr}$ decreases with time 
in Fig. \ref{fig:weakstrongB}.
Once $B_\mathrm{cr}$ becomes smaller than $B_0$, the magnetic field 
cannot be bent by the shock compression.
Magnetic tension realigns the magnetic field with the upstream field direction.
If the gravitational force is important, the gas accumulates along the realigned magnetic field.


Although all local simulations considered the MC formation via a single compression 
of the atomic gas, 
we need to consider multiple episodes of compressions to form 
giant MCs \citep{Inutsuka2015,Kobayashi2017,Kobayashi2018}.
Recent global simulations of galactic disks have also pointed out 
that global MCs experience collisional build-up driven by 
large-scale flows associated with the spiral arm and 
nearby supernova explosions during their growth and evolution \citep[e.g.,][]{Dobbs2015,Baba2017}.
We will investigate the effect of multiple compressions during the MC formation 
in a local simulation to achieve high resolution in the future.

\subsubsection{Limitation of the Analytic Model}
The analytic model developed in Sections \ref{sec:ana} and \ref{sec:parameter}
describes the evolution of the averaged quantities $\nsh$, $\Pmag$, and $\Pturb$ 
using the pressure balance across a post-shock layer.
However, the post-shock dynamics is not fully understood because 
we do not explain why $\delta v_x$ obeys the universal law shown in equation (\ref{dvx_para}) 
for $\theta>\thetacr$.
It also remains an unsettled question what determines $\thetacr$
although $\thetacr'$ is derived by the simple argument in Equation (\ref{th_critd}).
We expect that the interaction 
between the upstream CNM clumps and the warm interclump gas is crucial in the post-shock dynamics.
If $\theta$ is sufficiently small, the motion of CNM clumps triggers 
magnetic reconnection, which prevents the field lines from being compressed \citep{Jon1996}.
The interaction between CNM clumps through magnetic field lines 
also plays an important role in the post-shock dynamics \citep{Clifford1983,Elmegreen1988}.
We will address these issues in forthcoming papers.


\section{Summary}\label{sec:summary}
We investigated the early stage of the formation of MCs by colliding flows of the atomic gas,
which has a realistic two-phase structure where HI clouds are embedded in warm diffuse gases.
As parameters, we consider the mean density $\nave$, 
the strength and direction of the magnetic field $B_0$, and the collision speed $V_0$ 
of the atomic gas.

First, we investigated the MC formation at a fiducial parameter set of 
$(\nave=5~\pcc,V_0=20~\mathrm{km~s^{-1}},B_0=5~\mu\mathrm{G})$
by changing the angle $\theta$ between the magnetic field and upstream flow.
We focus on super-Alfv{\'e}nic colliding flows $(V_0>B_0/\sqrt{4\pi\rhoave})$ which 
are more likely for compression of denser atomic gases (Section \ref{sec:sub}).
Our findings are as follows.

\begin{enumerate}
     \item In the early phase, shock compression of the highly inhomogeneous atomic gas drives
           the velocity dispersions in the compression direction,
           which are as large as $\sim 0.3V_0$, regardless of $\theta$. 
           The transverse velocity dispersions are much smaller than the longitudinal ones.

   \item The later time evolution of the post-shock layers can be classified in terms of
           a critical angle $\thetacr$ which is roughly equal to $\sim 11^\circ$ 
           for the fiducial parameter set.

         \item If $\theta<\thetacr$, 
          an upstream CNM clump is not decelerated when it passes through a shock front, and 
          finally collides with the shock front on the opposite side.
          A highly super-Alfv{\'e}nic velocity dispersion is 
          maintained by accretion of the upstream CNM clumps.
          The velocity dispersion is highly biased in the compression direction.
          The magnetic field does not play an important role, and it is passively bent
          and stretched by the super-Alfv{\'e}nic gas motion.
          The post-shock layer significantly
  expands in the compression direction as a result of the large velocity dispersion.

    \item If $\theta>\thetacr$,
          the shock-amplified transverse magnetic field decelerates the CNM clumps moving in
          the compression direction. The decelerated 
	       CNM clumps are accumulated in the central region. 
          The velocity dispersion in the compression direction 
          decreases as $\propto t^{-1/2}$, and appears not to depend on $\theta$ sensitively.  
          The time dependence may be explained if there is a mechanism that keeps 
          the total kinetic energy in the compression direction constant. 
          The velocity dispersion in the transverse direction does not decrease with time.

   \item As a function of $\theta$, 
           the mean post-shock densities have a sharp peak 
           at an angle near $\theta\sim\thetacr$, regardless of time (Fig. \ref{fig:denave}).
         Around $\theta\sim \thetacr$, 
         the gas compression occurs not only in the collision direction 
         but also along the shock-amplified transverse magnetic field.
         As a result, the total mass of dense gases rapidly increases and 
         CO molecules efficiently form even without self-gravity.
         The mean post-shock densities, the dense gas masses, and the CO abundances 
         rapidly decrease toward $\theta<\thetacr$ ($\theta>\thetacr$)
         because of the ram pressure (magnetic stress).

  \item By developing an analytic model and 
          performing a parameter survey in the parameters of ($\nave,V_0,B_0$), 
        we derive an analytic formula for the critical angle $\theta_\mathrm{cr}$ 
        (Equation (\ref{th_crit})), above which 
        the shock-amplified magnetic field controls the post-shock dynamics.
        We also found that the mean ram and magnetic pressures in the post-shock layers 
        evolve in a universal manner as a function of 
	${\cal M}_\mathrm{A\perp}$ and the accumulated column density for various 
	parameter sets ($\nave,V_0,B_0$).  The evolution of the mean post-shock densities 
	also follows a universal law.  However, in colliding flows with 
        Alfv{\'e}n Mach numbers less than $\sim 4$, lower collision speeds, and 
        higher mean upstream densities, the post-shock mean densities can be high for $\theta<\thetacr$.

  \item The critical angle $\theta_\mathrm{cr}$ takes a small value 
        as long as magnetic fields are not very weak ($>1~\mu$G) (Fig. \ref{fig:thcrit}).
        If the atomic gas is compressed from 
	  various directions with respect to the field line direction, 
        the compression with $\theta<\thetacr$ seems to be rare.
        Although we need further simulations including self-gravity, 
        shock compression with angles a few times larger than the critical angle is expected not to 
        contribute to an efficient MC formation 
        because the CO formation is inefficient owing to the lack of dense gases.

\end{enumerate}
Our results show that 
the post-shock structures depend strongly on $\theta$ 
in the case that $\theta$ is less than a few times $\thetacr$.
This may create a diversity of the physical properties of dense clumps and cores in 
simulations including self-gravity.
Self-gravity is also expected to affect the global structures of MCs.
We will address the effect of self-gravity on the MC formation in forthcoming papers.

\section*{Acknowledgements}
We thank the anonymous referee for many constructive comments.
Numerical computations were carried out on Cray XC30 and XC50 at the CfCA of the 
National Astronomical Observatory of Japan and at the 
Yukawa Institute Computer Facility.
This work was supported in part by the Ministry of Education, Culture, Sports, Science and
Technology (MEXT), Grants-in-Aid for Scientific Research, 16H05998 (K.T. and K.I.), 16K13786 (K.T.), 
15K05039 (T.I.), 16H02160, 16F16024, 18H05436, and 18H05437 (S.I.).
This research was also supported by MEXT as “Exploratory Challenge
on Post-K computer” (Elucidation of the Birth of Exoplanets [Second Earth] and the Environmental Variations of
Planets in the Solar System).
\software{Athena++ \citep[][in preparation]{Stone2018}}

\appendix
\section{The Effect of Radiative Cooling on the Evolution of CNM Clumps}\label{app:cnm}

The interaction of a shock wave with an isolated interstellar cloud for the 
adiabatic case was investigated by \citet{KMC1994}. 
According to their estimation, the cloud is destroyed and mixed with the surrounding diffuse gas 
by the Kelvin-Helmholtz and Rayleigh-Taylor instabilities 
in several cloud-crushing times given by 
\begin{equation}
     t_\mathrm{cc,CNM} = \chi^{1/2} \frac{L_\mathrm{CNM}}{V_0} 
     \sim 
     0.2~\mathrm{Myr}\left( \frac{\chi}{ 20 } \right)^{1/2}
     \left( \frac{L_\mathrm{CNM}}{1~\mathrm{pc}} \right)
     \left( \frac{V_0}{20~\mathrm{km~s^{-1}}} \right)^{-1},
     \label{tcc}
\end{equation}
where $\chi$ is the density contrast between the CNM and WNM, and 
$t_\mathrm{cc,CNM}$ is the timescale for the shock to travel the cloud size 
in the compression direction, $L_\mathrm{CNM}$.
If there is turbulence in the post-shock region, the destruction timescale becomes even 
shorter \citep{Pittard2009}.
Thus, shocked CNM clumps are expected to be destroyed quickly on a timescale less than 1~Myr.

\begin{figure}[htpb]
     \centering
     \includegraphics[width=8cm]{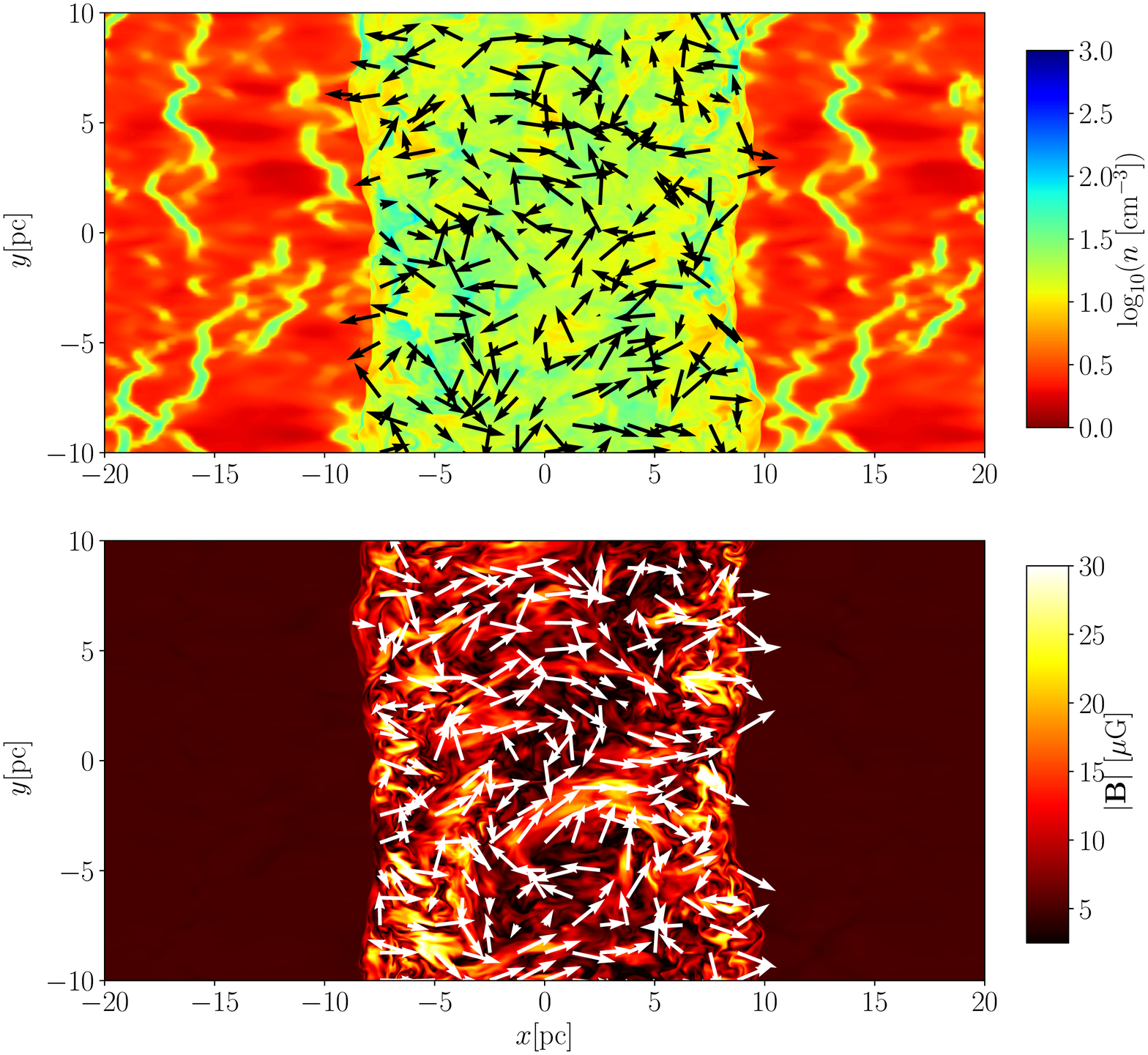}
     \caption{
     Maps of density (top panel) and field strength (bottom panel) in the $z=0$ plane at $t=1.2$~Myr for the adiabatic case 
     where the cooling/heating processes are switched off in the same initial and boundary conditions.
     The arrows in the top and bottom panels indicate the directions of the velocity and magnetic fields, respectively.
     }
     \label{fig:slice_ad}
\end{figure}

For comparison, we perform another simulation without 
the cooling/heating processes for the $\theta=0^\circ$ case.
The results are shown in Fig. \ref{fig:slice_ad}.
After passing through the shock fronts, the CNM clumps are quickly destroyed and 
are mixed up with the surrounding warm gases.
The mixing  causes 
the density distribution to be relatively smooth as shown in 
the top panel of Fig. \ref{fig:slice_ad}.
Unlike the results with radiative cooling,
the velocity field in the post-shock layer appears not to be biased in the compression direction 
but to be randomized.
This is consistent with the estimation in equation \ref{tcc}.
The bottom panel of Fig. \ref{fig:slice_ad} shows local amplification of the magnetic field.
This is formed by the Richtmyer-Meshkov instability, where 
the MHD interaction between a shock wave and a CNM clump
develops vortices that amplify magnetic fields \citep{InoueYamazaki2009,Sano2012}.

Fig. \ref{fig:3d}, however, shows that the two-phase structure is clearly seen in 
the post-shock layers for all the models, indicating that the cooling/heating processes are 
important in the survival of the CNM clumps.
This effect has been pointed out by \citet{Mellema2002}, \citet{Melioli2005},
and \citet{Cooper2009} in different contexts.
The significance of the radiative processes is characterized by 
the cooling time given by $P/\left\{ (\gamma-1)\Lambda \right\}$,
where we use the following approximate cooling rate:
\begin{equation}
  \Lambda \sim 3\times 10^{-28} n^2 \sqrt{T} e^{-92/T}~\mathrm{erg~cm^{-3}~s^{-1}}
  \label{lam}
\end{equation}
\citep{KI2002}. 
Just behind the shock front, 
the CNM temperature increases up to $3\mu m_\mathrm{H}V_0^2/16k_\mathrm{B}$ and 
the CNM density increases by a factor of 4 if the shock is adiabatic.
The cooling time of the shocked CNM is estimated as
\begin{equation}
  t_\mathrm{cool,CNM} \sim 
  1.2\times 10^{-2}~\mathrm{Myr} \left( \frac{V_0}{20~\mathrm{km~s^{-1}}} \right)
  \left( \frac{n_\mathrm{CNM,sh}}{200~\mathrm{cm}^{-3}} \right)^{-1},
  \label{tc}
\end{equation}
where $n_\mathrm{CNM,sh}$ is a typical CNM density in the post-shock region under the adiabatic assumption.
Equations (\ref{tcc}) and (\ref{tc}) indicate that the cooling time is much shorter than 
the cloud-crushing time.
A shocked CNM clump quickly cools and condenses to reach a thermal equilibrium state.
Although the CNM clumps fragment, this does not 
lead to complete mixing of
the CNM clumps with the surrounding WNM.
Instead, gas exchange owing to phase transition between CNM/WNM occurs \citep{II2014,Valdivia2016}.
The large density contrast also contributes to 
the long lifetime of the CNM clumps (see Equation (\ref{tcc})).
Magnetic fields also increase the lifetime of the CNM clumps by reducing 
the growth rate of the Kelvin-Helmholtz instability.

\section{
Evolution of the Hydrogen Molecule Fraction in a Post-shock Layer.
}\label{app:H2}

In this appendix, we estimate the H$_2$ fraction in a post-shock layer with 
a constant density of $n_\mathrm{sh}$.
Since the fresh material, whose H$_2$ fraction is almost zero, is continuously 
supplied to the post-shock layer, the evolution of the H$_2$ fraction in the post-shock layer 
cannot be understood using a simple formation time of $(k_\mathrm{H_2}n_\mathrm{sh})^{-1}$,
where $k_\mathrm{H_2}$ is the H$_2$ formation rate coefficient \citep{HM1979}.

Here we derive the H$_2$ fraction in the post-shock layer at an epoch of $t=t_\mathrm{f}$ as follows.
The time evolution of the number density of atomic H is given by 
\begin{equation}
        \frac{dn_\mathrm{H}}{dt} = -2 k_\mathrm{H_2} n_\mathrm{sh} n_\mathrm{H}.
        \label{H2eq}
\end{equation}
In order to derive an upper limit on the H$_2$ abundance, we neglect the photodissociation 
of H$_2$.
The number of the hydrogen nuclei entering the post-shock layer from $t$  to $t+\Delta t$ 
is $2L_0^2V_0 \nave \Delta t$.
Among them,
the number of the hydrogen nuclei remaining in atomic H is given by 
\begin{equation}
        \Delta N_\mathrm{H} = 2L_0^2V_0\nave \Delta t \exp\left[ -\int_{t}^{t_\mathrm{f}}
        2k_\mathrm{H_2}n_\mathrm{sh} dt\right],
        \label{NH}
\end{equation}
where we use the fact that 
most of the hydrogen nuclei are in atomic H just after the shock compression.
By integrating Equation (\ref{NH}) from $t=0$ to $t=t_\mathrm{f}$, one 
obtains the total number of atomic H at $t=t_\mathrm{f}$ as follows:
\begin{equation}
        N_\mathrm{H} = 2L_0^2 V_0 \nave \int_0^{t_\mathrm{f}} \exp\left[ -\int_{t}^{t_\mathrm{f}}
        2k_\mathrm{H_2}n_\mathrm{sh} ds\right]dt.
        \label{NH1}
\end{equation}
If $n_\mathrm{sh}$ and $k_\mathrm{H_2}$ are constant with time, Equation (\ref{NH}) is reduced to 
\begin{equation}
        N_\mathrm{H} = 2L_0^2 V_0 \nave \frac{1-e^{-2k_\mathrm{H_2} n_\mathrm{sh}t_\mathrm{f}}}
        { 2k_\mathrm{H_2} n_\mathrm{sh}}.
\end{equation}
Finally, the H$_2$ fraction $f_\mathrm{H2,ave}$ is given by 
\begin{equation}
        f_\mathrm{H2,ave} \equiv 
        \frac{N-N_\mathrm{H}}{N} = 1- \frac{1-e^{-2k_\mathrm{H_2} n_\mathrm{sh} t_\mathrm{f}}}
        { 2k_\mathrm{H_2} n_\mathrm{sh}t_\mathrm{f}},
\end{equation}
where $N=2L_0^2V_0\nave t_\mathrm{f}$ is the total number of the hydrogen nuclei 
accumulated in the post-shock layer until $t=t_\mathrm{f}$.

\section{The Effect of Self-gravity 
in the Case that Magnetic Fields are Almost Parallel to Colliding Flows.
}\label{app:grav}
Here we estimate the timescale on which a freely propagating CNM clump 
leaving the shocked region is decelerated by the gravitational force from the shocked region.
Let us consider a sheet with a surface density $\sigma$. If it is formed by the converging flows, 
we can write $\sigma=2\rhoave V_0 t$.  Since the gravitational force exerted 
by the sheet on an object with mass $m$ from the sheet is given by $F_\mathrm{G}=2\pi m G \sigma$, the equation of motion 
for a cold clump that is freely flying outside of the sheet is given by $dv/dt=2\pi G \sigma$.  
Thus, a clump that is ejected from the sheet with a velocity $v_\mathrm{ej}$ is attracted by 
the self-gravitational force of the sheet and 
eventually pulled back at the time $t_\mathrm{g}\simeq \{v_\mathrm{ej}/(4\pi G \rhoave V_0)\}^{1/2}$.  
If we use typical parameters of the $\theta=3^{\circ}$ case, the timescale can be expressed as
\begin{equation}\label{eq:tg}
t_\mathrm{g}\simeq 5~\mathrm{Myr}~
\left(\frac{v_\mathrm{ej}}{5~\mathrm{km~s}^{-1}}\right)^{1/2} 
\left(\frac{\nave}{5~\mathrm{cm}^{-3}}\right)^{-1/2}
\left(\frac{V_0}{20~\mathrm{km~s^{-1}}}\right)^{-1/2}.
\end{equation}
This suggests that the shocked region would not expand after $t_\mathrm{g}\simeq 5$ Myr.


\listofchanges

\end{document}